\documentclass[aps,pra,superscriptaddress,floatfix,amsmath,amssymb,reprint]{revtex4-2}
\pdfoutput=1

\usepackage{graphicx}

\usepackage{graphicx}
\usepackage{enumitem}
\usepackage{bbm}
\usepackage{relsize}
\usepackage{xcolor}
\usepackage[caption=false]{subfig}
\usepackage{amsmath}	
\usepackage{amssymb}    
\usepackage[pdftex]{hyperref}
\begin{document}




\title{Plunging in the Dirac sea using graphene quantum dots: there and back again}

\author{Fran\c{c}ois Fillion-Gourdeau}
\email{francois.fillion@emt.inrs.ca}
\affiliation{Institute for Quantum Computing, University of Waterloo, Waterloo, Ontario, Canada, N2L 3G1}
\affiliation{Infinite Potential Laboratories, Waterloo, Ontario, Canada, N2L 0A9}

\author{Pierre Levesque}
\affiliation{Infinite Potential Laboratories, Waterloo, Ontario, Canada, N2L 0A9}


\author{Steve MacLean}
\email{steve.maclean@emt.inrs.ca}

\affiliation{Institute for Quantum Computing, University of Waterloo, Waterloo, Ontario, Canada, N2L 3G1}
\affiliation{Infinite Potential Laboratories, Waterloo, Ontario, Canada, N2L 0A9}
\affiliation{Universit\'{e} du Qu\'{e}bec, INRS-\'{E}nergie, Mat\'{e}riaux et T\'{e}l\'{e}communications, Varennes, Qu\'{e}bec, Canada J3X 1S2}

\date{\today}

\begin{abstract}
The dynamics of low energy charge carriers in a graphene quantum dot subjected to a time-dependent local field is investigated numerically. In particular, we study a configuration where a Coulomb electric field is provided by an ion traversing the graphene sample. A Galerkin-like numerical scheme is introduced to solve the massless Dirac equation describing charge carriers subjected to space- and time-dependent electromagnetic potentials and is used to evaluate the field induced interband transitions. It is demonstrated that as the ion goes through graphene, electron-hole pairs are generated dynamically via the adiabatic pair creation mechanism around avoided crossings, similar to electron-positron pair generation in low energy heavy ion collisions. 
\end{abstract}


\maketitle


\section{Introduction}

In the last two decades, Dirac materials have received an unprecedented amount of attention because they have very interesting electrical, mechanical and thermal properties \cite{dirac_materials}. Graphene in particular, a 2-D array of carbon atoms arranged on a honeycomb lattice,  is the quintessential Dirac material. Close to its Dirac points at a momentum $|\boldsymbol{K}_{\pm}| \approx 1.7 \;\mbox{\AA}^{-1} \approx 12.3 \; \mbox{eV}/v_{F}$, the dispersion relation is linear and relativistic-like, being given by $E = v_{F}|\boldsymbol{p}| + O(|\boldsymbol{p}|/|\boldsymbol{K}_{\pm}|)$, where $v_{F}\approx 1.093 \times 10^{6}$ m/s is the Fermi velocity and $\boldsymbol{p}$ is the momentum deviation from $\boldsymbol{K}_{\pm}$ \cite{PhysRevB.76.081406}. In addition, the dynamics of charge carriers are described by a massless 2-D Dirac equation, as long as $|\boldsymbol{p}| \ll |\boldsymbol{K}_{\pm}|$, confirming the Dirac material nature of graphene \cite{RevModPhys.81.109}. 

It was realized early that owing to this relativistic-like quantum behavior, the dynamics of charge carriers in graphene would be similar to the one of relativistic electrons and thus, could be used to simulate or investigate quantum electrodynamics (QED) processes \cite{katsnelson2007graphene,gusynin2007ac}. The Klein paradox \cite{katsnelson2006chiral,PhysRevLett.102.026807}, the Zitterbewegung \cite{Zawadzki_2011}, atomic collapse \cite{PhysRevLett.99.246802, Wang734},  the Schwinger process \cite{PhysRevD.78.096009,PhysRevB.81.165431,zubkov2012,PhysRevLett.102.106802,PhysRevD.87.125011,PhysRevB.92.035401,PhysRevD.86.125022,akal2016low} and the Breit-Wheeler process \cite{golub2019dimensionality} have been considered from this point of view. Therefore, whilst there may exists some qualitative difference between graphene and QED \cite{gusynin2007ac}, this material provides a link between condensed matter and high energy physics.   

\begin{figure}
	\begin{center}
		\includegraphics[width=0.40\textwidth]{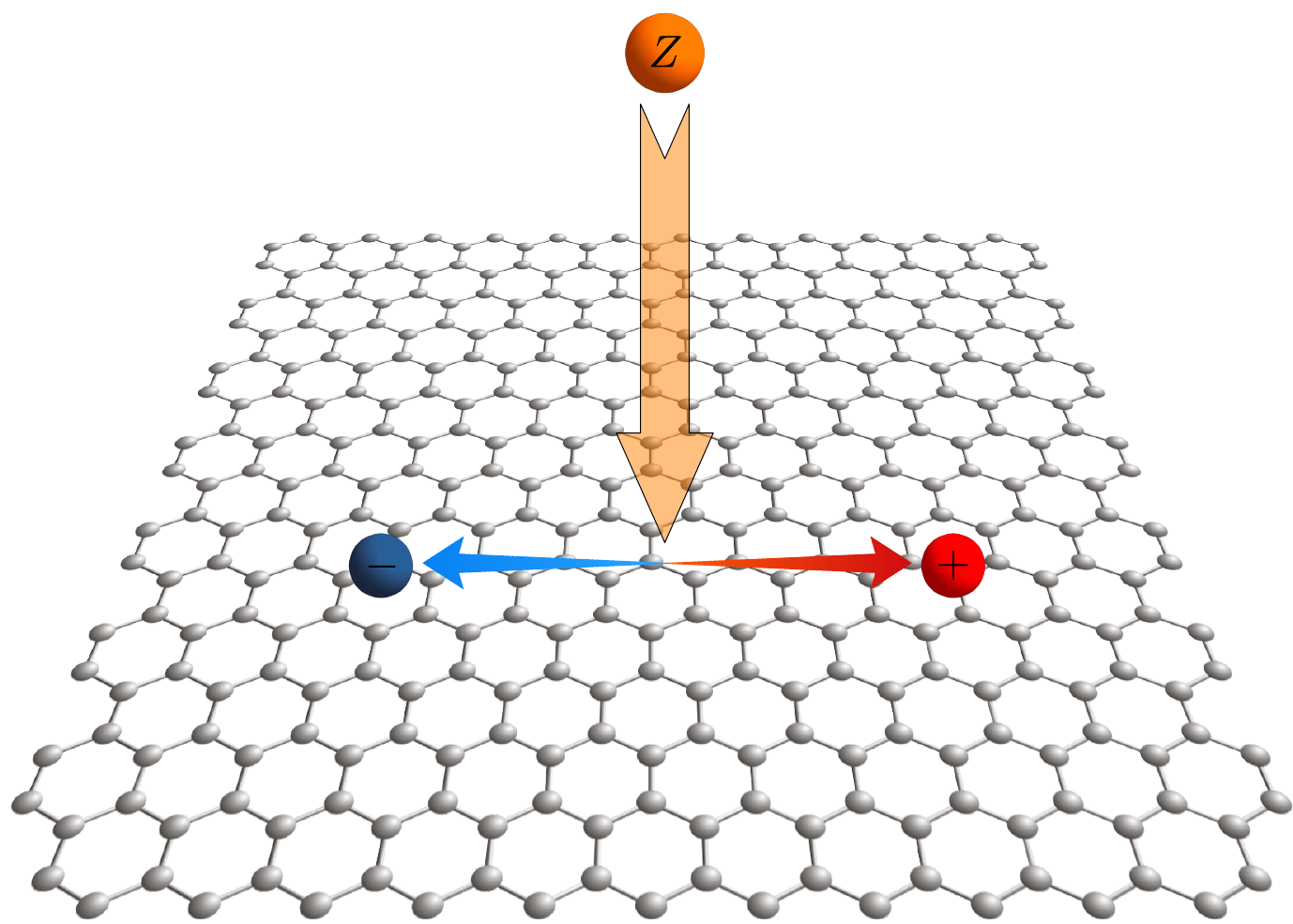}
	\end{center}
	\caption{Ion bombarded graphene where an ion of charge $Z$ passes through a graphene sample and generates electron-hole pairs. }
	\label{fig:ion_bombarded_graphene}
\end{figure}

In this article, electron-hole pair production in ion bombarded graphene is proposed as an analogue to QED electron-positron pair production in low energy heavy ion collisions (HIC), in the same spirit as these previous studies. The configuration considered is shown in Fig. \ref{fig:ion_bombarded_graphene}. As the ion traverses the finite graphene sample (graphene quantum dot), it interacts with charge carriers in the valence band and has a certain probability to excite them in the conduction band, thus generating an electron-hole pair. In the following, we argue that under certain conditions, pair production occurs via the same mechanism as in HIC whereby some positive energy states ``plunge'' into the negative energy continuum (the Dirac sea).

The dynamics of charge carriers in graphene subjected to electromagnetic fields has gained momentum in the last few years, with the potential of controlling the electron dynamics in graphene-based devices \cite{Gagnon:18,Lefebvre:18,higuchi2017light,PhysRevLett.121.207401}. Many theoretical and experimental investigations have focused on the interaction of graphene with a (homogeneous) laser field  \cite{PhysRevB.84.205445,PhysRevB.83.153410,PhysRevB.84.205406,PhysRevB.85.125413,PhysRevB.88.035430}. The regime of strong fields, whereby multi-photonic effects, the holy grail of atomic physics, start to be important and may lead to new physical phenomena, have also been investigated \cite{PhysRevB.91.045439,akal2016low,PhysRevB.92.035401,PhysRevD.86.125022,PhysRevD.87.125011,PhysRevLett.102.106802,Al_Naib_2015,PhysRevLett.109.166603,Heide_2019}. For inhomogeneous fields, some work has also been performed. For example, some experiments \cite{COMPAGNINI20093201} and numerical simulations using either time-dependent density functional theory (TDDFT) simulations \cite{PhysRevB.81.153401,PhysRevB.85.235435,0953-8984-27-2-025401,PhysRevB.89.035120}, nonequilibrium Green functions methods \cite{PhysRevB.94.245118} and molecular dynamics \cite{doi:10.1021/jp208049t}, have investigated ion bombardment of graphene samples. The  main objectives of these studies was to quantify the amount of energy transferred from the projectile to the target, the stopping power, and understand the formation of defects in the atomic structure as the charge traverses the honeycomb lattice. Another example is Ref. \cite{gruber2016ultrafast}, where the current generated by highly charged ions (with $Z=20-30$, where $Z$ is the ion electric charge) have been estimated experimentally and using DFT calculations. The spatio-temporal dynamics of charge carriers under local optical excitation has been considered in Ref. \cite{jago2019spatio}. Nonetheless, the dynamics of electrons and holes subjected to localized inhomogeneous fields is not as well known as in the homogeneous case. One of the goals of this article is to fill this gap by investigating electron-hole pair production from a dynamical and localized Coulomb-like potential. 

To study the main features of this dynamical and non-perturbative phenomenon, extensive numerical simulations are performed to evaluate the pair production rate and the electron density generated by the passing ion. Assuming that certain conditions (given below) are fulfilled, the charge carriers are modeled by a simple Dirac equation, allowing for a connection with QED processes and HIC. To solve this space- and time-dependent Dirac equation, a Galerkin-like numerical scheme is introduced, similar to the ones developed in Refs. \cite{0953-4075-37-20-005,PhysRevA.91.032708,FILLIONGOURDEAU2016122}. Finally, we discuss the challenges for observing this phenomenon experimentally.    

This article is separated as follows. In section \ref{sec:pp_prod}, the main process for particle-hole production in graphene and its analogy with electron-positron pair creation are presented. Section \ref{sec:model} is devoted to the physical model for charge carriers. The numerical method is introduced in Section \ref{sec:num_meth} while numerical results can be found in Section \ref{sec:num_res}. We conclude in Section \ref{sec:conclu}. Natural units where $\hbar = c =1$ are used throughout the article.

\section{Plunging in the Dirac sea: graphene and HIC \label{sec:pp_prod}} 

In the Dirac sea interpretation, pair production is generated by an external field that induces transitions between the negative and positive energy states \cite{Greiner:1985,Rafelski2017}. In condensed matter systems, this corresponds to interband transitions between the valence and conduction bands. In the second-quantization formalism, these transitions between the negative and the positive continua are actually responsible for the generation of electron-hole or electron-positron pairs \cite{Greiner:1985}. 

\begin{figure}
	\begin{center}
		\includegraphics[width=0.40\textwidth]{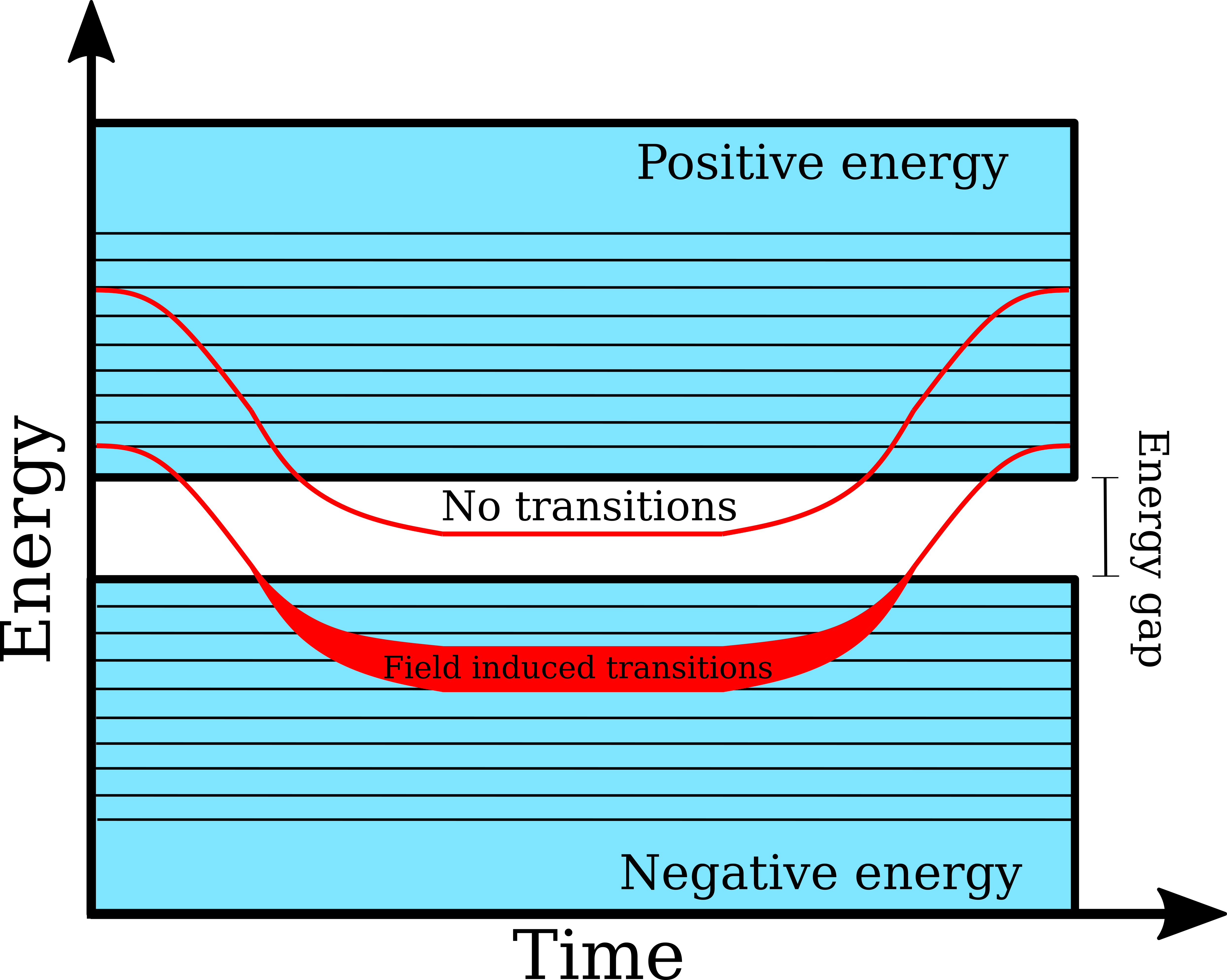}
	\end{center}
	\caption{Single-particle spectrum of the Dirac operator for pair production via the APC mechanism. The blue color represents the possible free positive and negative energy states separated by an energy gap $\Delta$, while the small black lines are bound states. In red are two specific states. The bottom one is shifted as the ion approaches each other, falls into the Dirac sea and crosses with negative energy states. As this occurs, it becomes unstable, its spectral width increases and transitions are possible. The top one is also shifted by the potential but the shift is not large enough to generate transitions.  }
	\label{fig:plunging}
\end{figure}

More specifically, the mechanism considered in this article is adiabatic pair creation (APC) \cite{Pickl_2008} (sometimes called spontaneous pair creation \cite{Rafelski2017}), a dynamical QED process where the external electromagnetic field potential has some specific characteristics: 
\begin{enumerate}
	\item Time-dependent potential well that can localize the electron (form bound states or resonances) at some given time,
	\item Supercritical for some time interval, where the positive bound states energies dive into the negative energy continuum, 
	\item Adiabatic and non-perturbative time evolution.  
\end{enumerate}
The single-particle state of electrons subjected to external electromagnetic fields of this type is depicted in Fig. \ref{fig:plunging}. This figure also includes an energy gap, which is induced by the finite size of the graphene sample (see the end of Sec. \ref{sec:time_ind_free}). The lowest energy states of the positive energy continuum (valence band) are shifted towards the negative energy states (conduction band), or Dirac sea.  When electronic states dive into the Dirac sea, they cross with negative energy states and interband transitions can occur, resulting in the production of electron-hole pairs. Because initially, the negative energy states are filled (Dirac sea or valence band), the transitions always occur from the negative to the positive energy states. This physical interpretation is supported by the second-quantized formulation, as shown in Sec. \ref{sec:model}. 

In QED, the Coulomb field becomes supercritical when $Z \sim 137$. Ions with such high charges do not exist in nature, so the strategy used in HIC is to collide two highly charged ions (typically fully stripped Uranium atoms with $Z=92$). As the two ions approach each other, the bound state energies of the combined system decrease, up to a critical distance (at approximately $R_{\mathrm{cr}}\sim 35$ fm \cite{Rafelski2017}) where the lower energy state reaches $E = -mc^{2}$ and where the field becomes supercritical. At this point, the lower bound state starts ``plunging'' in the Dirac sea, i.e. it enters the negative energy continuum. This allows for transitions between negative and positive energy states and thus, allows for the generation of electron-positron pairs. After the collision, the bound states return to the positive energy continuum. Although this pair creation mechanism has been predicted in the 70's \cite{gershtein1970positron,muller1976positron}, it still eludes an experimental observation \cite{PhysRevLett.75.2658,RUFFINI20101,Rafelski2017}, in part because the ion ``sticking'' time is very short yielding a small number of pairs. 
 
In the considered configuration (see Fig. \ref{fig:ion_bombarded_graphene}), the field is provided by an ion passing through a graphene sample. As the ion passes, the lower energy states of the conduction band are pulled down and fall within the valence band, allowing for interband transitions. This occurs when $\beta > \frac{1}{2}$, i.e. for charges of value $Z \sim 1$ \cite{PhysRevLett.99.246802,PhysRevLett.99.166802,PhysRevB.80.165429}. Finally, as the ion gets further, we recover the free dynamics but electron-hole pairs have been created. This phenomenon is the dynamical analog of the atomic collapse resonances observed experimentally from artificial nuclei on graphene \cite{Wang734}.

The main differences between pair creation in HIC and electron-hole creation in ion bombarded graphene is the presence of the mass gap in HIC, which effectively reduces the pair production rate because it entails higher ion charges for the occurrence of pair production, and the absence of Coulombic bound states before the ion interaction. In other words, in graphene, the transitions occurs between scattering states. Nevertheless, the physical principle underlying electron-hole production in ion bombarded graphene and electron-positron pair production are the same: they are produced via APC. To verify if this phenomenon can be measured experimentally, we study the spectrum of the system and perform a numerical study where the electron-hole pair density is evaluated.    

\section{Graphene model and observables \label{sec:model}}

In this article, we consider a regime where the charge transfer from graphene to the ion is minimal and where the integrity of the graphene sample is preserved. According to some experiments \cite{gruber2016ultrafast,buchheim2016understanding} and time-dependent density functional calculations \cite{PhysRevB.85.235435,PhysRevB.81.153401}, this occurs for ion energy of 1 keV to 2 MeV and higher, and low charge number $Z \sim 1-2$. In these conditions, the number of defects and transferred electrons induced by the interaction of the projectile with graphene are negligible.

Provided that these conditions are fulfilled, we choose a theoretical description of charge carriers in terms of a massless Dirac equation where the speed of light $c$ is replaced by the Fermi velocity $v_{F} \approx 0.003 c$ \cite{RevModPhys.81.109,gusynin2007ac}. This description is accurate as long as the energy of the charge carriers is $E \lesssim 2$ eV, where corrections due to the tight-binding model can be neglected \cite{RevModPhys.81.109}. Throughout this article, we are assuming that these conditions are fulfilled. It will be verified \textit{a posteriori} that most electrons are generated with an energy lower than 2 eV. In addition, the interactions between electrons and the environment (through phonon-electron and electron-electron interactions) should be negligible. This can be controlled to a certain extent by using a high dielectric constant substrate and a very low temperature setting, which ensures a small electron-electron coupling constant $\alpha_{G} \sim e^{2}/4\pi \epsilon v_{F} \lesssim 1$, where $\epsilon$ is the dielectric constant of the substrate. This condition also allows us to neglect recombination of carriers into photons, a process $\propto \alpha_{G}$ which should not modify the electron distribution significantly for the electronic densities reached in the system.
Finally, the dynamics should occur on a time scale smaller than the thermalization time, on the order of a few tens of femtoseconds \cite{PhysRevB.83.153410}. Otherwise, the energy distribution of the charge carriers will be simply given by a thermal Fermi-Dirac-like distributions where the features of the dynamics have been washed out.

Considering an interaction of graphene with a charged particle, there is an azimuthal symmetry around the collision point. Then, it is possible and convenient to express the Dirac equation in polar coordinates:
\begin{align}
\label{eq:dirac_eq}
i\partial_{t} \psi(t,r) =  \left\{- i v_{F}\begin{bmatrix}
0 & \partial_{r} + \frac{\mu_{2}}{r}\\
\partial_{r} - \frac{\mu_{1}}{r} & 0
\end{bmatrix}  + V(t,r)\right\}\psi(t,r) , 
\end{align}
where $\psi(t,r)$ is the two-component wave function, $r$ is the radial distance, $V$ is an angle-independent scalar potential, $\mu_{1,2} = j_{z} \mp 1/2$ and $j_{z}$ is the $z$-angular momentum quantum number taking half-integer values ($j_{z} = \cdots,-\frac{3}{2},-\frac{1}{2},\frac{1}{2},\frac{3}{2},\cdots$). 

As we are interested into the collision of an ion with a graphene sample, the scalar potential represents the field of the moving ion with velocity $v$ and charge $Z$. It is chosen as
\begin{align}
V(t,r) =
\begin{cases}
 -\cfrac{\alpha Z}{R(t)} & \mbox{for}\; R(t)>R_{\mathrm{reg}} \\
-\cfrac{\alpha Z}{2R_{\mathrm{reg}}}   \left(3 - \cfrac{R^{2}(t)}{R^{2}_{\mathrm{reg}}} \right) & \mbox{for}\; R(t)<R_{\mathrm{reg}}
\end{cases},
\end{align} 
where $\alpha \approx 1/137$ is the fine structure constant and $R(t):=\sqrt{(vt)^{2} + r^{2}}$ is the distance from the charge. The singular Coulomb potential is regularized by a spherical constant charge distribution for $R(t) \leq R_{\mathrm{reg}}$, where $R_{\mathrm{reg}}$ is the charge radius. The ion radius will be approximately the size of the graphene lattice constant $R_{\mathrm{reg}} \sim a \approx 0.246$ nm, similar to Ref. \cite{PhysRevB.80.165429}.   

The first observable considered in this article is the electron distribution on single-particle states (labeled by the energy $E_{k}$ and the angular momentum $j_{z}$), defined as
\begin{align}
n_{k j_{z}} = N_{\mathrm{v}}N_{\mathrm{spin}} \langle 0_{\mathrm{in}} | \hat{a}_{j_{z},k}^{(\mathrm{out}) \dagger} \hat{a}_{j_{z},k}^{(\mathrm{out})} | 0_{\mathrm{in}} \rangle,
\end{align}
where the subscripts $\mathrm{in}$ ($\mathrm{out}$) means that the mathematical object is evaluated at time $t \rightarrow +\infty$ ($t \rightarrow -\infty$) and where $\hat{a}_{j_{z},k}^{(\mathrm{out}) \dagger} ,\hat{a}_{j_{z},k}^{(\mathrm{out})}$ are creation/annihilation operators in the electron-hole representation for the single particle free electron states $u_{k j_{z}}(r)$ (states in the conduction band). We also introduced $N_{\mathrm{v}}=2$, the number of Dirac valleys, and $N_{\mathrm{spin}}=2$, the number of physical electron spin, to account for degeneracies.  

To take into account the finite nature of the sample, the electronic states are defined on a compact support $r \in [0,r_{\mathrm max}]$ with a boxed boundary condition, where one of the spinor components is set to zero at $r=r_{\mathrm{max}}$, that is 
\begin{align}
\begin{cases}
\psi_{1}(t,r_{\mathrm{max}}) = 0 & \mbox{for} \;\; j_{z} > 0 \\
\psi_{2}(t,r_{\mathrm{max}}) = 0 & \mbox{for} \;\; j_{z} < 0
\end{cases}.
\end{align}
Physically, this corresponds to the zigzag boundary condition, which has been used numerous times to model circular graphene quantum dots \cite{PhysRevB.77.035316,PhysRevB.84.205441}. With these boundary conditions, the free states in both conduction and valence bands are discrete ($k \in \mathbb{N}$) because electrons are confined. Other boundary conditions could also be considered like the armchair or the MIT bag model, but this is outside the scope of this article, which focuses on the dynamics of charge carriers.

Using the Furry picture, the electron distribution generated by the strong external field can be calculated from \cite{Greiner:1985}:
\begin{align}
\label{eq:av_num}
	n_{k j_{z}}  &= N_{\mathrm{v}}N_{\mathrm{spin}}\sum_{k'} \left|U_{j_{z},k,k'}(t_{f},t_{i}) \right|^{2},
\end{align}
where $t_{f},t_{i}$ are the final and initial time, respectively (outside of this time interval, the external field is turned off). We also defined the inner product 
\begin{align}
\label{eq:projection}
U_{j_{z},k,k'}(t_{f},t_{i}) &:= \langle u_{k j_{z}} | \psi_{k',j_{z}}(t_{f}) \rangle \\
&=   \int_{0}^{r_{\mathrm{max}}} u_{k j_{z}}^{ \dagger}(r)  \psi_{k',j_{z}}(r,t_{f}) rdr,
\end{align}
where the time-dependent wave function $\psi_{k j_{z}}(r,t)$ is a solution to the Dirac equation \eqref{eq:dirac_eq} with an initial condition given by a negative eigenstate: $\psi_{k j_{z}}(r,t_{i}) = v_{k j_{z}}(r)$ at initial time $t_{i}$, where the field is turned on. Also, we are assuming a null chemical potential ($\mu=0$) to simulate QED-like process, but this could be relaxed if one is interested in other graphene initial conditions. 

The second observable considered is the spatial electron density $\rho(r)$, evaluated from 
\begin{align}
\label{eq:density}
\rho_{j_{z}}(r) &= N_{\mathrm{v}}N_{\mathrm{spin}} \sum_{k'}\left|\sum_{k} U_{j_{z},k,k'}(t_{f},t_{i}) u_{k j_{z}}(r) \right|^{2}.
\end{align}
It can be verified that one recovers the electron distribution in Eq. \eqref{eq:av_num} by integrating the density on the domain and by using the orthogonality of the wave function $u_{k j_{z}}(r)$.

\section{Numerical method \label{sec:num_meth}}

To evaluate the observables, we need to determine the wave function $\psi_{k j_{z}}(r,t)$.
As long as $\alpha Z/v \gtrsim 1$, i.e. for a ratio of the ion charge and velocity which is small enough, a perturbative treatment for finding the wave function is not accurate: perturbation theory does not hold. Therefore, one should solve the Dirac equation Eq. \eqref{eq:dirac_eq} non-pertubatively. For this purpose, we employ a high order Galerkin numerical scheme. The rationale for this choice is threefold: first, it can deal with the polar coordinate singularity (in $1/r$) in a straightforward way, second, it is very efficient for the evaluation of time-independent states $u_{k j_{z}}(r)$,$v_{k j_{z}}(r)$ required in the calculation and finally, the order of convergence of the spatial discretization is high.

The numerical calculation proceeds in three distinct steps: 1) The time-independent free solutions $u_{k j_{z}}(r)$ and $v_{k j_{z}}(r)$ are determined, 2) The negative energy states are evolved according to the Dirac equation with the ion field and 3) The time-dependent wave function is projected onto positive energy state, using Eq. \eqref{eq:projection}. Once the function $U_{j_{z},k,k'}$ is determined, we have access to both observables.

\subsection{Time-independent scheme: Rayleigh-Ritz method \label{sec:time_ind}}

The time-independent Dirac equation is given by
\begin{align}
\label{eq:dirac_eq_ti}
E_{k} \phi_{k j_{z}}(r) =  \left\{- i v_{F}\begin{bmatrix}
0 & \partial_{r} + \frac{\mu_{2}}{r}\\
\partial_{r} - \frac{\mu_{1}}{r} & 0
\end{bmatrix}  + V_{t}(r)\right\}\phi_{k j_{z}}(r) , 
\end{align}
where $E_{k}$ is the $k$'th eigenenergy, $\phi_{k j_{z}}(r)$ is the $k$'th eigenstate and $V_{t}(r) = V(t,r)$ is the potential evaluated at some specific time. Two numerical schemes are introduced to solve this eigenvalue problem: one for the free case, when $V_{t}(r)=0$ and one for the interacting case, when $V_{t}(r) \neq 0$. The former is used to initialize the time-dependent solver while the latter is used to study the spectral characteristics of the system. 

\subsubsection{The free case: $V_{t}(r) = 0$ \label{sec:time_ind_free}}

In the free case with $V_{t}(r) = 0$, the analytical solution of Eq. \eqref{eq:dirac_eq_ti} is well-known (see Appendix \ref{app:free_sol}). In principle, it is possible to interpolate the solution with the B-spline basis set expansion given below by solving a linear system. However, it is more convenient, for the same computational complexity, to actually solve the eigenvalue problem.

When there is no potential, for $j_{z}>0$ (the case $j_{z}<0$ is discussed in Appendix \ref{app:neg_jz}), Eq. \eqref{eq:dirac_eq_ti} can be written for the first spinor component as 
\begin{align}
\label{eq:ti_u1}
\left(\partial_{r}^{2} + \frac{1}{r}\partial_{r} - \frac{\mu_{1}^{2}}{r^{2}} + \frac{E^{2}_{k}}{v_{F}^{2}} \right)u_{1,k j_{z}}(r)=0, \\
\label{eq:ti_u2}
u_{2, k j_{z}}(r) =  i \frac{v_{F}}{E_{k}} \left( \partial_{r} - \frac{\mu_{1}}{r} \right) u_{1, k j_{z}}(r), 
\end{align}
where we focus on the positive energy states by setting $\phi_{k j_{z}}(r) = u_{k j_{z}}(r)$. 
The negative energy solution can be calculated from the positive one as $u_{1, k j_{z}}(r) = v_{1, k j_{z}}(r)$ and $u_{2, k j_{z}}(r) = -v_{2, k j_{z}}(r)$. 
The form given in \eqref{eq:ti_u1}-\eqref{eq:ti_u2} is particularly convenient because the spectrum of Eq. \eqref{eq:ti_u1} is bounded from below, allowing for a variational Rayleigh-Ritz (RR) method. The latter is equivalent to minimizing the following  energy functional obtained from Eq. \eqref{eq:ti_u1} (here, we dropped the index $k j_{z}$ for convenience):
\begin{align}
\label{eq:functional}
\mathcal{E} &: = \lambda \int_{0}^{r_{\mathrm{max}}}rdr u_{1}^{*}(r)u_{1}(r) \nonumber \\
&- \int_{0}^{r_{\mathrm{max}}}rdr u_{1}^{*}(r) \left(\partial_{r}^{2} + \frac{1}{r}\partial_{r} - \frac{\mu_{1}^{2}}{r^{2}} \right)u_{1}(r),
\end{align}
where the eigenvalue is $\lambda = \frac{E^{2}_{k}}{v_{F}^{2}}$. This can be approximated by expanding the solution on a polynomial basis over the domain $[0,r_{\mathrm{max}}]$ as 
\begin{align}
\label{eq:basis_exp}
u_{1}(r) = r^{|\mu_{1}|} \sum_{i=1}^{N} a_{1,i} B_{i}(r),
\end{align}
where $\{a_{1,i}\}_{i=1,\cdots,N}$ are the basis coefficients, $\{B_{i}(r)\}_{i=1,\cdots,N}$ are the basis functions and the prefactor $r^{|\mu_{1}|}$ was added to facilitate the implementation of the boundary condition at $r=0$. Indeed, the behavior of the solution is given by $u_{1}(r) = r^{|\mu_{1}|} F(r^{2})$ \cite{FILLIONGOURDEAU2014559}, where the function can be Taylor expanded as $F(r^{2}) \sim A_{0} + A_{2}r^{2} + \cdots$, implying that $\partial_{r}F(r^{2})|_{r=0} = 0$. 

In this work, a B-spline polynomial basis set $\{b_{i}^{(p)}(r)\}_{i=1,\cdots,N_{s}}$, where $N_{s}$ is the number of B-splines and $p$ is the order of the spline polynomial, is chosen because B-splines $b_{i}^{(p)}(r)$ have compact support, allowing for numerical sparse matrix linear algebra packages \cite{nla.cat-vn991654}. Also, both the Dirac and the Schrodinger equation have been solved with high accuracy using these basis sets  \cite{0034-4885-64-12-205,0953-4075-37-20-005,JPSJ.75.114301,PhysRevA.37.307,0953-4075-42-5-055002}.
The B-spline set and the basis function set are related by $\{B_{i}(r)= b_{i+1}^{(p)}(r)\}_{i=2,\cdots,N}$. The first basis function $B_{1,1}(r)$ deserves a special treatment to take the boundary condition at $r=0$ into account. This boundary condition can actually be implemented by a suitable combination of B-spline functions, as $B_{1,1}(r) = b_{1}^{(p)}(r) + c b_{2}^{(p)}(r)$. The constant $c$ is adjusted such that $\partial_{r} B_{1}(r)|_{r=0} = 0$, forcing the exact boundary condition. 

B-splines are determined by the polynomial order $p$ and the knot vector (see \cite{0034-4885-64-12-205} for more details). In this work, an equidistant knot vector is used where each knot is separated by the element size $h$. In addition, the knot multiplicity is chosen in agreement with the implementation of the boundary conditions. In particular, the multiplicity is $p$ at the endpoint $r_{\rm{min}} = 0$, allowing for the canceling of the derivative of the first basis function. On the other hand, the multiplicity is $p-1$ at  $r_{\rm{max}}$ insuring a boxed boundary condition where $u_{1}(r_{\mathrm{max}}) = 0$. Finally, the multiplicity is 1 at the interior points, guaranteeing the continuity of the solution. This yields a knot vector given by $[r_{1}, \cdots, r_{2p+n_{\mathrm{b}}-3}]$, with knot points ordered as
\begin{align}
r_{\rm{min}}  = r_{1}  = \ldots & = r_{p} < r_{p+1} < \cdots   \nonumber \\ 
< r_{p+n_{\mathrm{b}}-1} & = \ldots =r_{2p+n_{\mathrm{b}}-3} = r_{\rm{max}} \, ,
\end{align}
where $\{r_{p+j} = jh\}_{j=1,\cdots n-1}$. Here, $n_{\mathrm{b}}$ is the number of breakpoints, related to the number of B-spline as $N_{s} = n_{\mathrm{b}}+p-3$.

An approximation of the eigenenergy and eigenvector is found by minimizing the functional $\mathcal{E}$ over the coefficients of the basis function expansion. This is performed as usual in the RR method by reporting the basis set Eq. \eqref{eq:basis_exp} into Eq. \eqref{eq:functional}. The latter becomes a generalized eigenvalue problem of the form
\begin{align}
\lambda A \boldsymbol{a} = D\boldsymbol{a},
\end{align}
where $\boldsymbol{a}$ is a vector with entries $\boldsymbol{a} = (a_{1,1},a_{1,2},\cdots,a_{1,N})^{T}$ while $A$ and $D$ are Hermitian matrices. The entries in these matrices essentially consists in integrals of basis functions (and their first derivative) over the simulation domain. They are given explicitly in Appendix \ref{app:mat_tide}. Because we are using B-splines for the basis functions, which have compact support, the matrices $A$ and $D$ are sparse. The integrals over basis functions are performed numerically using the Gauss-Legendre (GL) quadrature, while the generalized eigenvalue problem is solved using \textsc{LAPACK} \cite{laug}. The solution of the eigenvalue problem then yields the first spinor component of the positive and negative energy states ($u_{1}$ and $v_{1}$, respectively). The other component $u_{2}$ is then evaluated using Eq. \eqref{eq:ti_u2}. Therefore, the second spinor component is given by
\begin{align}
\label{eq:basis_exp_second}
u_{2}(r) =    r^{|\mu_{1}|} \sum_{i=1}^{N} a_{2,i} \partial_{r} B_{i}(r), 
\end{align}
where $a_{2,i} =  i \frac{v_{F}}{E}a_{1,i}$, which actually defines the basis function expansion for the second spinor component. This choice of basis expansion for $u_{2}$ guarantees that the relation between the two spinor components Eq. \eqref{eq:ti_u2} is fulfilled everywhere on the domain. However, it implies that both components are expressed using different basis functions, similar to techniques for the Dirac equation based on kinetically balanced basis expansion \cite{PhysRevLett.93.130405}. However, the latter is not required in the free and massless case because the spinor components can be decoupled.

It was verified in Appendix \ref{app:num_conv_ti} that this numerical scheme reproduces the eigenenergies and eigenstates of the free Dirac operator. The latter can be solved analytically, as shown in Appendix \ref{app:free_sol}. In both cases, one can note the presence of an energy gap between positive and negative energy states, given by $\Delta = 2 v_{F}j_{0,1}/r_{\mathrm{max}}$. This effective gap is induced by the finite size of the graphene sample and vanishes in the limit $r_{\mathrm{max}} \rightarrow \infty$.   

\subsubsection{The interacting case: $V(t,r) \neq 0$\label{sec:time_ind_int}}

In the presence of a potential, the two spinor components cannot be decoupled as in the free case. For this reason, a slightly different approach is used here, based on a direct discretization of Eq. \eqref{eq:dirac_eq_ti}. However, the spectrum of this equation is not bounded from below, in contrast to Eq. \eqref{eq:ti_u1}. In this case, the naive utilization of the Rayleigh-Ritz method can lead to spectral pollution, i.e. the appearance of spurious eigenstates \cite{lewin,QUA:QUA560250112}. This problem can be mitigated by using balanced basis functions \cite{PhysRevLett.93.130405,lewin} or different variational principles \cite{PhysRevLett.57.1091,lewin}. In this work, we are using kinetically balanced basis functions \cite{doi:10.1063/1.447865} using a similar strategy as given in Ref. \cite{PhysRevA.85.022506}.

From Eq. \eqref{eq:dirac_eq_ti}, we can obtain the following RR functional:
\begin{align}
\label{eq:functional_int}
\mathcal{E} &: = \int_{0}^{r_{\mathrm{max}}}rdr \biggl[|\phi_{1}(r)|^{2} + |\phi_{2}(r)|^{2} \biggr] (E-V_{t}(r))\nonumber \\
&-iv_{F} \int_{0}^{r_{\mathrm{max}}}rdr \biggl[ \phi_{1}^{*}(r) \left(\partial_{r} + \frac{\mu_{2}}{r} \right) \phi_{2}(r)\biggr] \nonumber \\
&-iv_{F} \int_{0}^{r_{\mathrm{max}}}rdr \biggl[ \phi_{2}^{*}(r) \left(\partial_{r} - \frac{\mu_{1}}{r} \right) \phi_{1}(r)\biggr].
\end{align}
In the kinetically basis set approach, the basis expansion is given by
\begin{align}
\label{eq:basis_exp_bal1}
\phi_{1}(r) &= r^{|\mu_{1}|} \sum_{i=1}^{N} a_{1,i} B_{i}(r),\\
\label{eq:basis_exp_bal2}
\phi_{2}(r) &= r^{|\mu_{1}|} \sum_{i=1}^{N} a_{2,i} \partial_{r} B_{i}(r).
\end{align}
Reporting this basis expansion in Eq. \eqref{eq:functional_int} yields the generalized eigenvalue problem
\begin{align}
\label{eq:ti_sys_int}
E S \boldsymbol{a}(t) = \left[ C + P \right]\boldsymbol{a}(t),
\end{align}
where $S,C,P$ are matrices similar to the ones in the time-independent case (their explicit expression is given in Appendix \ref{app:mat_tdde}). The vector is also different because it now contains both spinor components, ordered as $\boldsymbol{a}(t) = (a_{1,1}(t),a_{2,1}(t), \cdots, a_{1,N}(t), a_{2,N}(t))^{T}$. This marks an important difference with the free case, where the solution of the eigenvalue problem yields only the first spinor component while the second component can be calculated from Eq. \eqref{eq:basis_exp_second}. In the interacting case, the solution to Eq. \eqref{eq:ti_sys_int} gives both spinor components. The free case is slightly more efficient because the number of rows and columns of the matrices in the eigenvalue problem is half of the interacting case.

\subsection{Time-dependent scheme: Galerkin method \label{sec:time_dep}}


Adapting the time-independent solver to the time-dependent case is relatively straightforward by using a Galerkin method \cite{FILLIONGOURDEAU2016122}. The basis functions expansion has the same form as Eqs. \eqref{eq:basis_exp_bal1} and \eqref{eq:basis_exp_bal2}, but the coefficients become time-dependent as (for $j_{z}>0$) 
\begin{align}
\psi_{1}(t,r) &= r^{|\mu_{1}|}\sum_{i=1}^{N} a_{1,i}(t)B_{1,i}(r), \\
\psi_{2}(t,r) &= r^{|\mu_{1}|}\sum_{i=1}^{N} a_{2,i}(t) \partial_{r} B_{1,i}(r).
\end{align}
The time-dependent basis expansion is then substituted in the Dirac equation \eqref{eq:dirac_eq}. As usual, the Galerkin method is then obtained by projecting this equation on the set of all basis functions $\{(r^{|\mu_{1}|}B_{1,i},0),(0,r^{|\mu_{1}|}\partial_{r} B_{1,i})\}_{i=1,\cdots,N}$, which allows us to obtain an equation of the form
\begin{align}
\label{eq:td_sys}
iS \partial_{t} \boldsymbol{a}(t) = \left[ C + P(t) \right]\boldsymbol{a}(t),
\end{align}
where $S,C,P(t)$ are matrices similar to the ones in the time-independent case (their explicit expression is given in Appendix \ref{app:mat_tdde}). The vector contains both spinor components, ordered as $\boldsymbol{a}(t) = (a_{1,1}(t),a_{2,1}(t), \cdots, a_{1,N}(t), a_{2,N}(t))^{T}$. The initial condition is fixed to the solution obtained from the time-independent solver by setting $a_{a,i}(t_{i}) = a_{a,i}$, for all $i \in [1,N]$ and $a = 1,2$. 

Eq. \eqref{eq:td_sys} is a large system of ordinary differential equation which yields the time dependence of the basis coefficients. We solve this system using the implicit Crank-Nicolson method \cite{FILLIONGOURDEAU2016122} whereby each time iteration $\Delta t$ is obtained by solving the following linear system of equations:
\begin{align}
\left[ S + i \frac{\Delta t}{2} \left( C + P^{n+\frac{1}{2}}\right) \right] \boldsymbol{a}^{n+1} = \nonumber \\
\left[ S - i \frac{\Delta t}{2} \left(C + P^{n+\frac{1}{2}}\right) \right]\boldsymbol{a}^{n},
\end{align}
where we defined $\boldsymbol{a}^{n} = \boldsymbol{a}(t_{n})$ and $P^{n+\frac{1}{2}} = P(t_{n}+\Delta t/2)$, along with $t_{n} = t_{i} + n\Delta t$. The linear system is solved in parallel by using a generalized minimal residual Krylov method implemented in the \textsc{Petsc} high-performance library \cite{petsc-web-page}. This numerical scheme has a second order convergence in time.

It was verified that the numerical scheme reproduces the time evolution of a free state and converge towards the exact solution in Appendix \ref{app:conv_td}. This time evolution can be computed analytically and is given in Appendix \ref{app:free_sol}.

\section{Numerical results and discussion \label{sec:num_res}}

This section is devoted to the numerical results obtained from the numerical schemes described in the previous section for the field induced transitions in ion bombarded graphene quantum dots. The first set of numerical results is an analysis of the spectrum of the Coulomb potential, to demonstrate that the electron-hole creation proceeds via the APC mechanism for the parameters considered subsequently. Then, the actual dynamics of charge carriers subjected to the field of the passing ion is evaluated. 

\subsection{Spectral characteristics of the Coulomb potential and adiabatic evolution \label{sec:num_res_spec}}

To determine conditions for which electron-hole pair production occurs via the APC mechanism, we now perform a detailed numerical analysis of the Coulomb potential spectrum. As shown below, the eigenstates of the system provide significant details on the physical processes at play. These results will be important in the next section, for the interpretation of the more complex dynamical results.

\begin{figure}
	\begin{center}
		\subfloat[]{\includegraphics[width=0.45\textwidth]{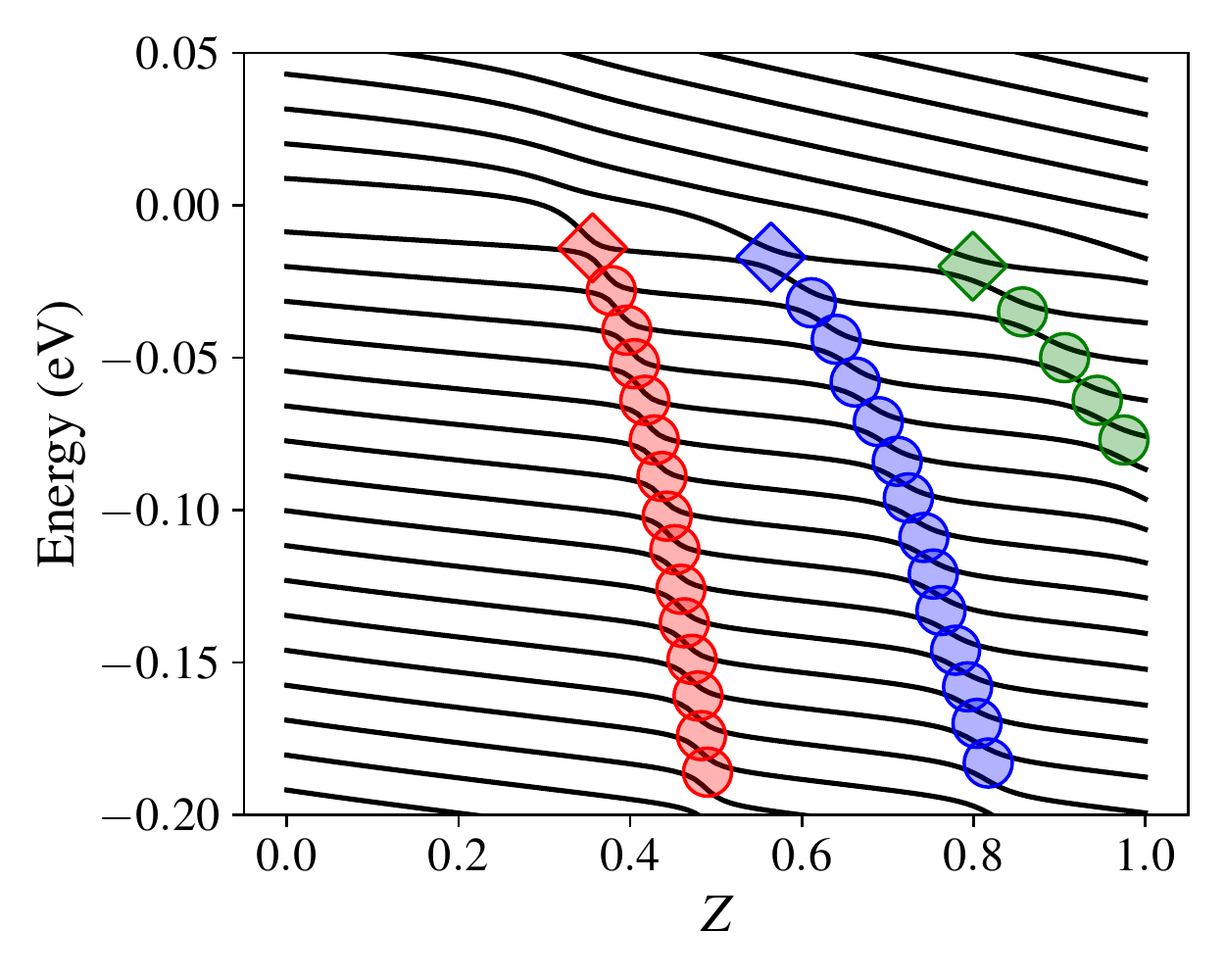}} \\
		\subfloat[]{\includegraphics[width=0.45\textwidth]{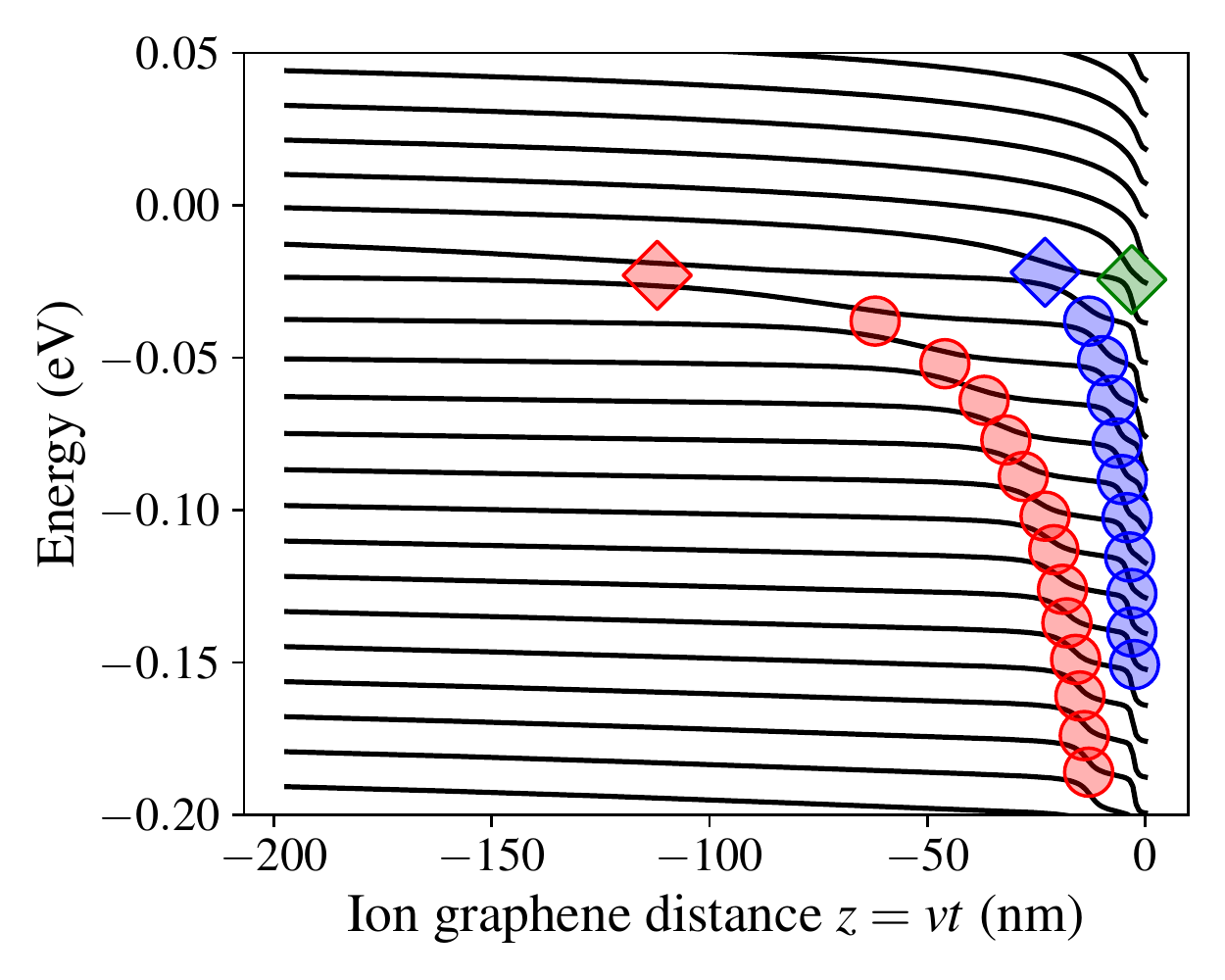}}
	\end{center}
	\caption{Eigenstates as a function of (a) ion charge $Z$ and (b) ion-graphene distance $z=vt$. The states of the conduction band (positive energy states) have energies $E_{k}>0$ when $Z=0$ while the states of the valance band (negative energy states) have $E_{k}<0$ when $Z=0$. The red diamond highlights the avoided crossing of the ground state with the first negative energy state while red circles are associated with avoided crossings of other negative energy states. The blue (green) diamonds and circles are highlighting avoided crossings for the first (second) positive excited states with negative energy states.  }
	\label{fig:eigen_dep}
\end{figure}

The eigenvalues of the Coulomb potential are obtained numerically by solving the time-independent Dirac equation \eqref{eq:dirac_eq_ti} using the numerical method described in Section \ref{sec:time_ind_int}. Two main parameters of the model are varied: the ion charge $Z$ and the graphene-charge distance $z$. The other simulation parameters are set to values given in Table \ref{tab:param}, the same values used for evaluating the charge carrier dynamics. For the case where $Z$ is varied, we fix the charge on the graphene sample by setting $z=vt =0$. The numerical results are displayed in Fig. \ref{fig:eigen_dep}.

As the ion charge increases, the positive energy states, states of the conduction band with energies $E_{k} >0$ for $Z=0$, are shifted towards lower energies. Remarkably, there is an avoided crossing between the positive energy state with the lowest energy (PSLE) and the negative energy state with the highest energy (NSHE) when the ion charge reaches a critical value $Z_{\mathrm{crit}} \approx 0.36$ (this is surrounded by a red diamond in Fig. \ref{fig:eigen_dep}-(a)), indicating that the Coulomb potential becomes supercritical and that electron-hole pairs can be created. Indeed, around this exceptional point, states involved in the avoided crossing become resonances (they actually acquire an imaginary part and cross in the energy imaginary plane \cite{Heiss_2012,Heiss_1990,PhysRevE.61.929}), suggesting that transitions between negative and positive energy states can occur via a non-perturbative tunneling process, where the transition probability in the adiabatic regime is given by the Landau-Zener formula \cite{wittig2005landau,joye1994proof}. These transitions between negative and positive energy states result physically in the creation of electron-hole pairs, as will be shown in the next section. As the ion charge is increased further, other avoided crossings occur between different energy states. These crossings are highlighted by red circles in Fig. \ref{fig:eigen_dep} for the PSLE and by blue and green diamonds and circles for the first and second states above PSLE, respectively. Each of these crossings corresponds to different potential tunnelling and pair creation pathways. 

These first numerical results demonstrate that the potential can become supercritical for some (low) ion charge value, fulfilling one of the criteria of APC. However, in ion bombarded graphene, the charge is not sitting on the graphene sample. To verify if avoided crossings also occur as the ion moves away from the graphene sample, the spectrum is evaluated as a function of the ion-graphene distance $z = vt$, for an ion charge $Z=1$. The results are displayed in Fig. \ref{fig:eigen_dep}-(b), using the same simulation parameters as when the ion charge is varied. These results show again avoided crossings between positive and negative energy states, highlighted by diamonds and circles in the figure. The first avoided crossing of the PSLE occurs at a critical time $t_{\mathrm{crit}} \approx -0.11$ $\mu$m$/v$, where $v$ is the ion velocity, implying that in the time interval $[-t_{\mathrm{crit}},t_{\mathrm{crit}}]$, the potential becomes supercritical, in agreement with the APC mechanism. Other states also participate and have avoided crossings with negative energy states, allowing for other tunneling pathways. 

\begin{figure}[t!]
	\begin{center}
		\includegraphics[width=0.40\textwidth]{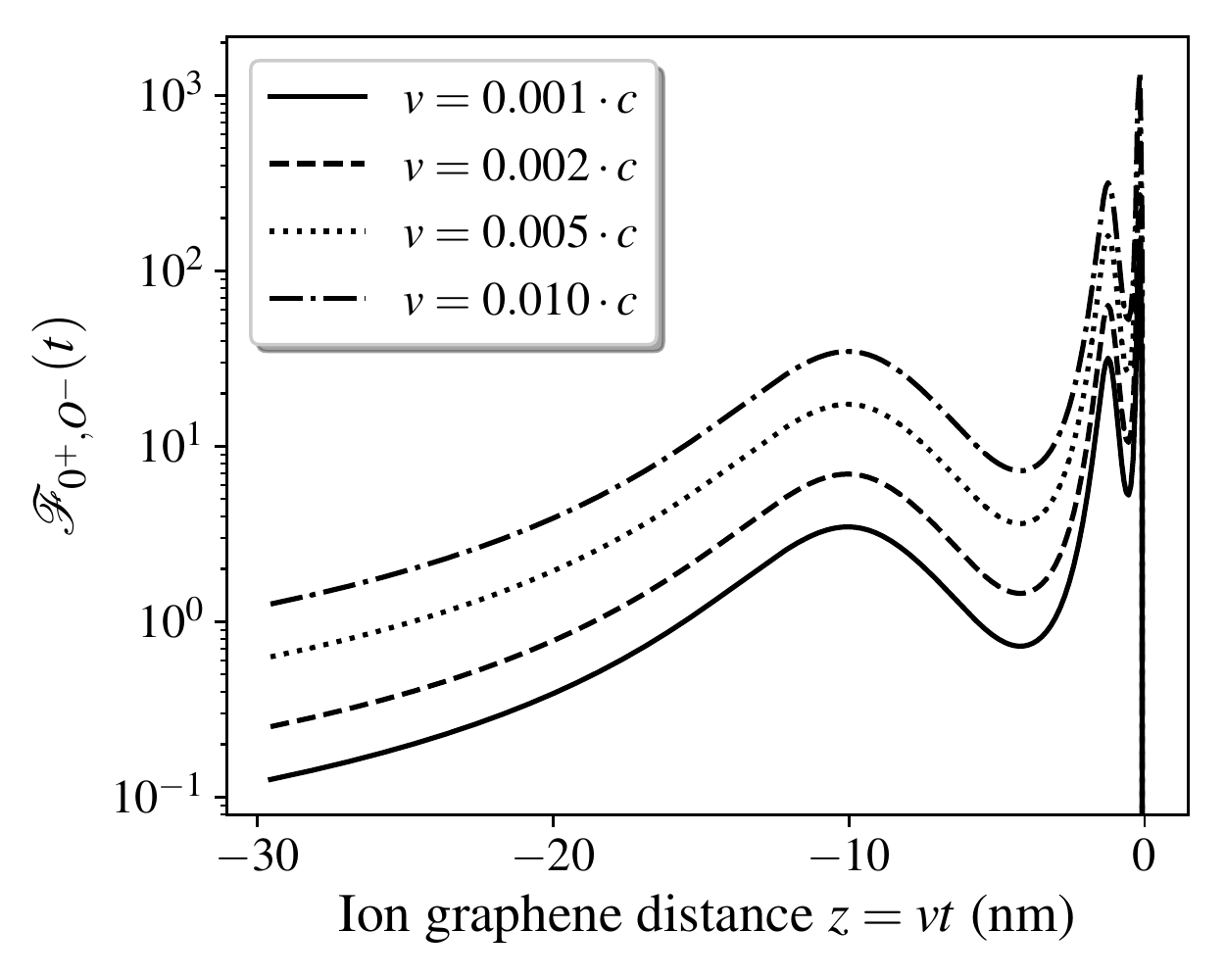}
	\end{center}
	\caption{Adiabatic parameters as function of ion graphene distance, for many velocities.}
	\label{fig:adiabat}
\end{figure}

These numerical results demonstrate that the first two criteria of APC given in Sec. \ref{sec:pp_prod} are fulfilled by the physical system under consideration: the potential can localize the electron by forming bound states and it can become supercritical in some time interval. The last criterion for APC is an adiabatic time evolution of the system along some particular times where non-adiabatic transitions occur. From adiabatic perturbation theory, a condition for adiabaticity can be found. It is expressed in terms of the adiabatic factor as \cite{sarandy2004consistency}
\begin{align}
\mathcal{F}_{kk'}(t) = \left| \frac{\langle \phi_{kj_{z}} | \partial_{t} V_{t}(r) |\phi_{k'j_{z} } \rangle }{\left[E_{k}(t) - E_{k'}(t) \right]^{2}} \right| \ll 1,
\end{align}   
for a transition between two eigenstates $k$ and $k'$.
This condition is fulfilled, when either the time evolution of the potential (and eigenstates) is slow or when the energy gap between the eigenstates is large. Both conspire to make the system adiabatic, whereby the transition rate between eigenstates $k$ and $k'$ is negligible. As an illustration, the adiabatic parameter is evaluated numerically and shown in Fig. \ref{fig:adiabat} for the most important transition of our system, i.e. between PSLE and NSHE (labeled as $0^{+}$ and $0^{-}$, respectively). This figure shows clearly that the adiabatic parameter is $\mathcal{F}_{0^{+},0^{-}}(t) > 1$ at avoided crossings (for $z \approx -0.01$ $\mu$m) while it obeys $\mathcal{F}_{0^{+},0^{-}}(t) \ll 1$ far from avoided crossings. This naturally leads to the following picture for the dynamics of interband field-induced transitions. When the ion is far from the graphene sample, the dynamics of charge carriers is adiabatic, no transition takes place and the eigenstates accumulate a phase. Then at some critical time (or distance), there is a non-adiabatic transition where the electron-hole pair creation process proceeds via tunnelling, as anticipated from the APC mechanism. The resulting electron distribution will be evaluated in the next section. We note that this process is actually very similar to field induced electronic transitions in molecules \cite{bandrauk1993molecules}. 

As expected, the adiabatic parameter reaches much higher values as the ion velocity is increased. However, the ion will spend less time around avoided crossings, resulting in a lower tunneling rate in the non-adiabatic transition. Therefore, the picture based on a sequence of adiabatic evolution followed by non-adiabatic transitions may start to fail at high enough $v$.

\subsection{Field induced interband transitions \label{sec:field_induced}}

The results presented in the last section showed that in some regime, electron-hole pair creation proceeds via APC. In this section, we are interested in the actual dynamics of charge carriers and distribution when they are subjected to the Coulomb potential of a passing ion. In short, we consider the dynamical production of electron-hole pair in ion bombarded graphene quantum dots, the configuration of Fig. \ref{fig:ion_bombarded_graphene}. The main observables are the electron distribution and the electron density, given in Eqs. \eqref{eq:av_num} and \eqref{eq:density}, respectively. They are calculated using numerical solutions of the Dirac equation using the Galerkin method described in Section \ref{sec:time_dep}. The value of important simulation parameters are summarized in Table \ref{tab:param}. It was tested empirically that converged results are obtained with these parameters by varying the time step and the number of basis functions. More details on the convergence of the numerical scheme can be found in Appendix \ref{app:num_conv}.  Only the first half of all calculated eigenstates are evolved in time because the error on higher energy states is too large and yields inaccurate results (see Fig. \ref{fig:error_vs_eigen} and the discussion in Appendix \ref{app:num_conv_ti}).  

\begin{table}[t!]
	\begin{tabular}{lc}
		\hline \hline
		Parameters & Value \\
		\hline
		Number of basis function ($N$) & 1024 \\
		B-spline order ($p$) & 4 \\
		Number of GL points & 10 \\
		Domain size ($r_{\mathrm{max}}$) & 0.197 $\mu$m \\
		Charge size ($R_{\mathrm{reg}}$) & 0.197 nm\\
		Initial time ($t_{i}$) & -65.8 fs \\
		Final time ($t_{f}$) & 65.8 fs \\
		Time increment ($dt$) & 0.013 fs\\
		\hline \hline
	\end{tabular}
	\caption{Simulation parameters for field induced transitions.}
	\label{tab:param}
\end{table}

For every numerical results presented in this section, a sum over angular momentum $j_{z}$ is performed. It was verified that for $|j_{z}|> 1/2$, all the contribution can be neglected because it is at least an order of magnitude below the ones for $|j_{z}| = 1/2$.

\begin{figure}[t!]
	\begin{center}
		\subfloat[]{\includegraphics[width=0.40\textwidth]{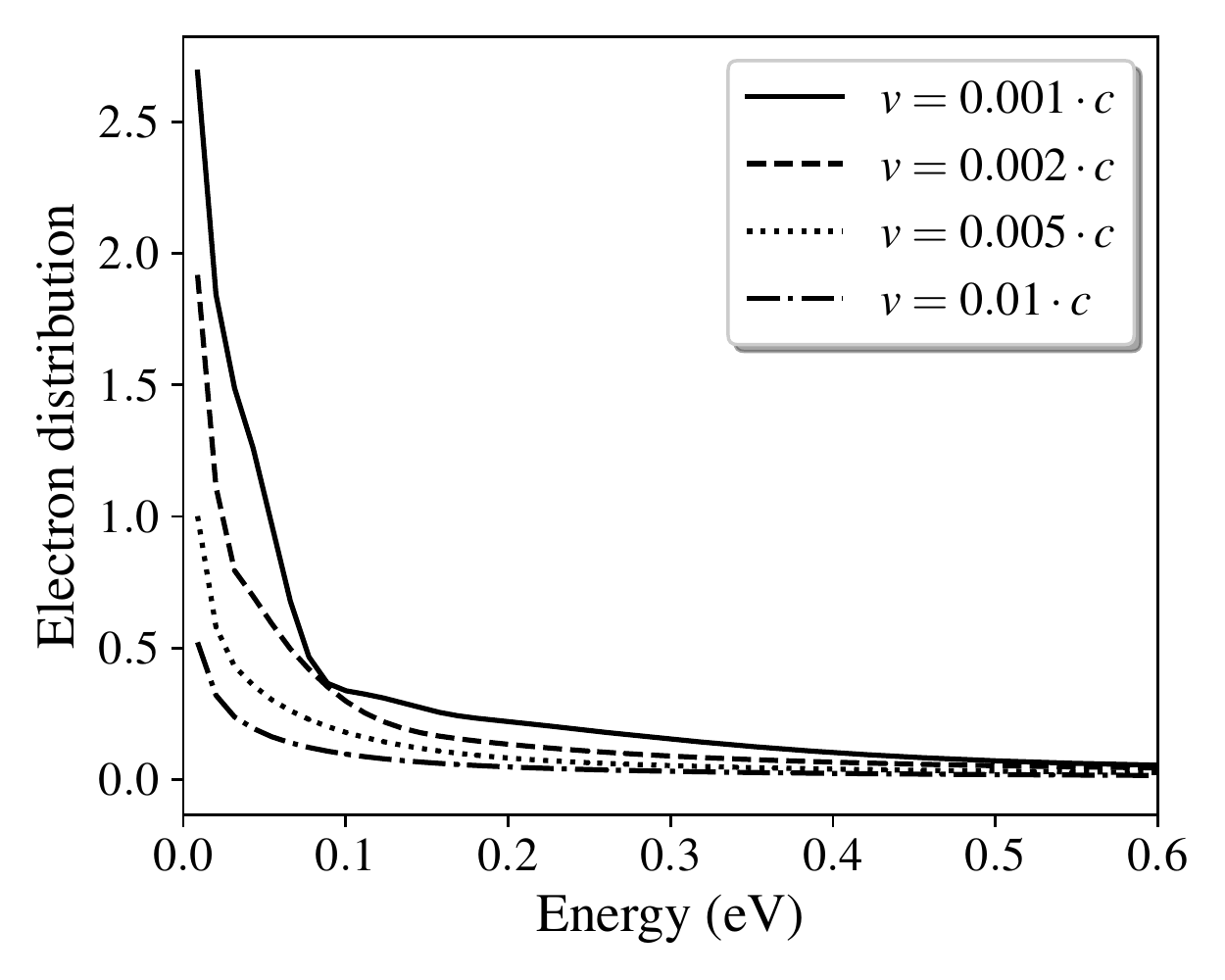}} \\
		\subfloat[]{\includegraphics[width=0.40\textwidth]{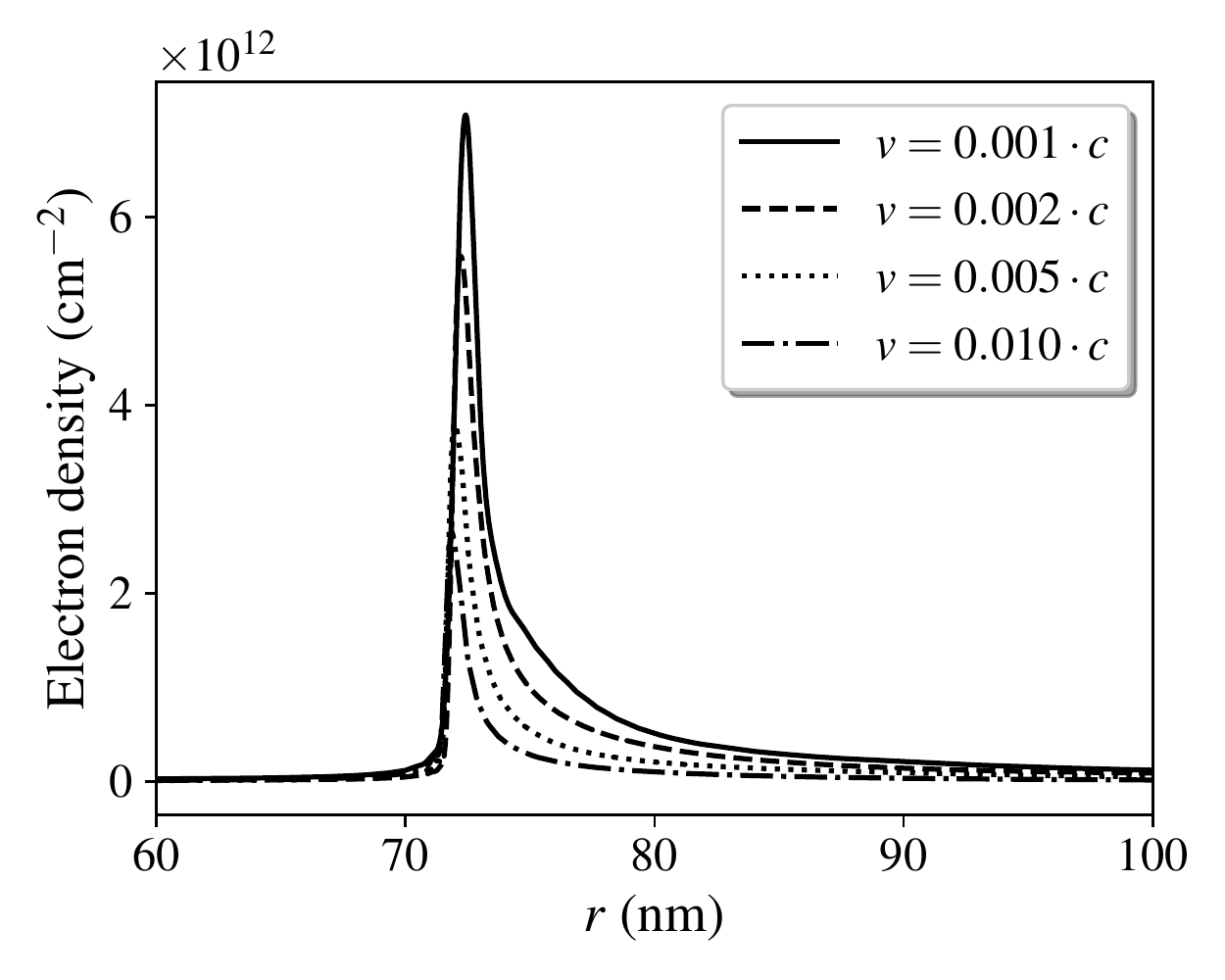}}
	\end{center}
	\caption{Electron distribution (a) and electron spatial density (b) as a function of the ion velocity $v$. The spatial density is shown at time $t=65.8$ fs. }
	\label{fig:vel_dep}
\end{figure}

The numerical results for the electron distribution and density are displayed in Fig. \ref{fig:vel_dep} for an ion charge $Z=1$ and for different velocities. The numerical results for the electron distribution in Fig. \ref{fig:vel_dep}-(a) demonstrate that most electrons are created at low energies, lower than 2 eV where the Dirac model is still a valid approximation. It was demonstrated in numerical studies of HIC that the lower energy electron peak is one of the qualitative features hinting at a supercritical potential \cite{PhysRevA.91.032708}. In addition, it was shown in the last section that most avoided crossings occur between low energy states, so the main contribution of these transitions should be found at low energies. For these reasons, the low energy electron peak suggests that electron-hole pairs are produced by the APC mechanism.  

Another feature of the electron distribution is that more electrons are generated at lower velocities. As argued in the last section, the adiabaticity of the system becomes dubious at higher velocities. Moreover, the faster ions stay for a shorter time in the vicinity of avoided crossings, resulting in a lower number of generated pairs through tunneling. From these considerations, one concludes that lower velocities are better suited for the investigation of APC. 

The spatial density at a specific final time $t = 65.8$ fs is displayed in Fig. \ref{fig:vel_dep}-(b). Qualitatively, the charge distribution is a ring centered around the ion impact point. The density peak evolves with time to larger radial distance $r$ (not shown here for simplicity). As it evolves, the density is reduced by a simple geometrical effect whereby the ring radius becomes larger and the density covers a larger area. In short, these results demonstrate that electrons are created in the neighborhood of the impact point, when the ion is close to the graphene sample. In the first few instants, the excitation is (quasi-)bound to the ion and localized around $r=0$. When the ion gets further from graphene, the Coulomb force is no longer strong enough to localize the charge carrier, so the latter is released and propagates freely in the radial direction, resulting in a radial current. The ion velocity does not modify this picture significantly, except for the fact that higher densities are reached at lower velocity, as expected from the results for the electron distribution.

\begin{figure}[t!]
	\begin{center}
		\subfloat[]{\includegraphics[width=0.40\textwidth]{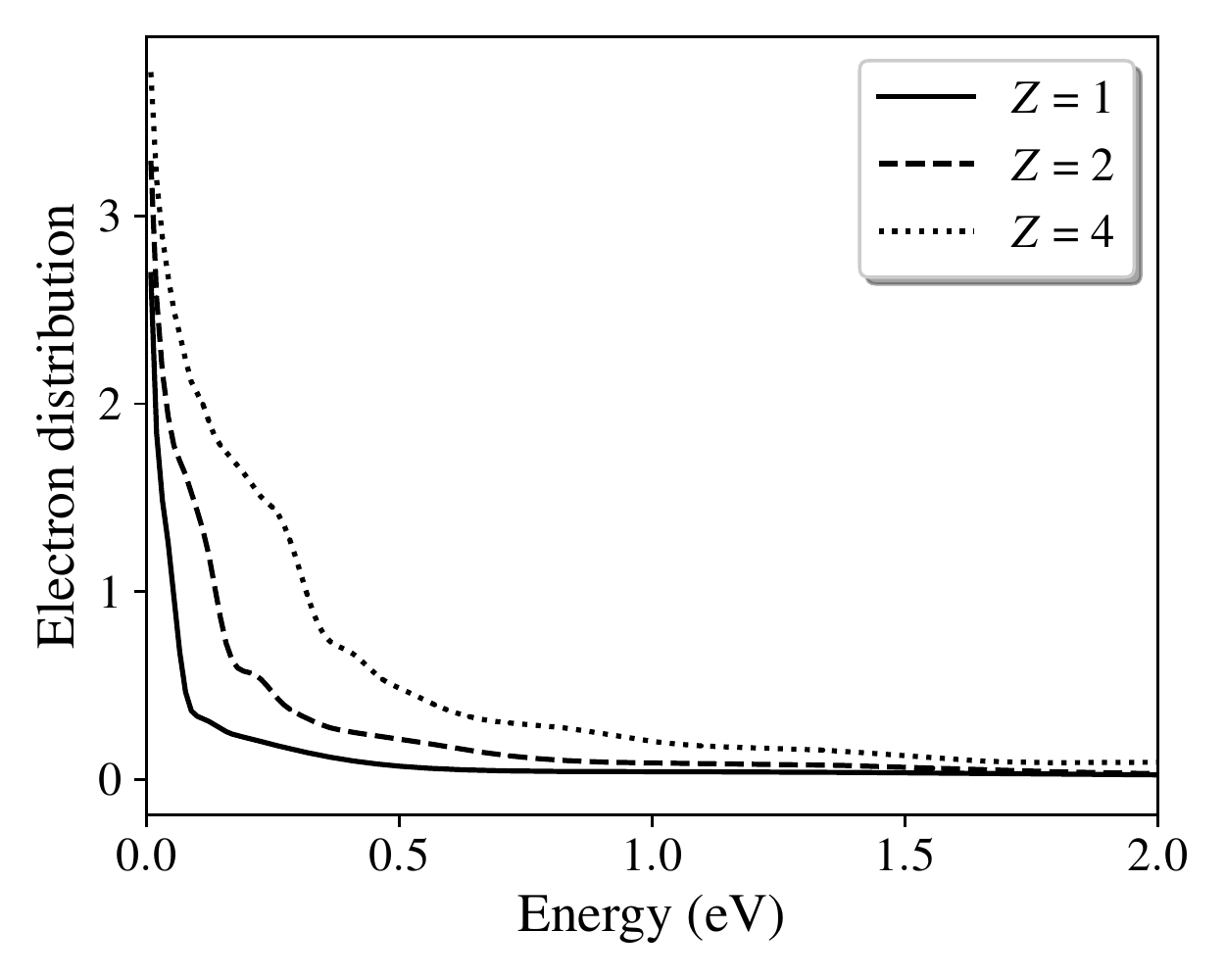}} \\
		\subfloat[]{\includegraphics[width=0.40\textwidth]{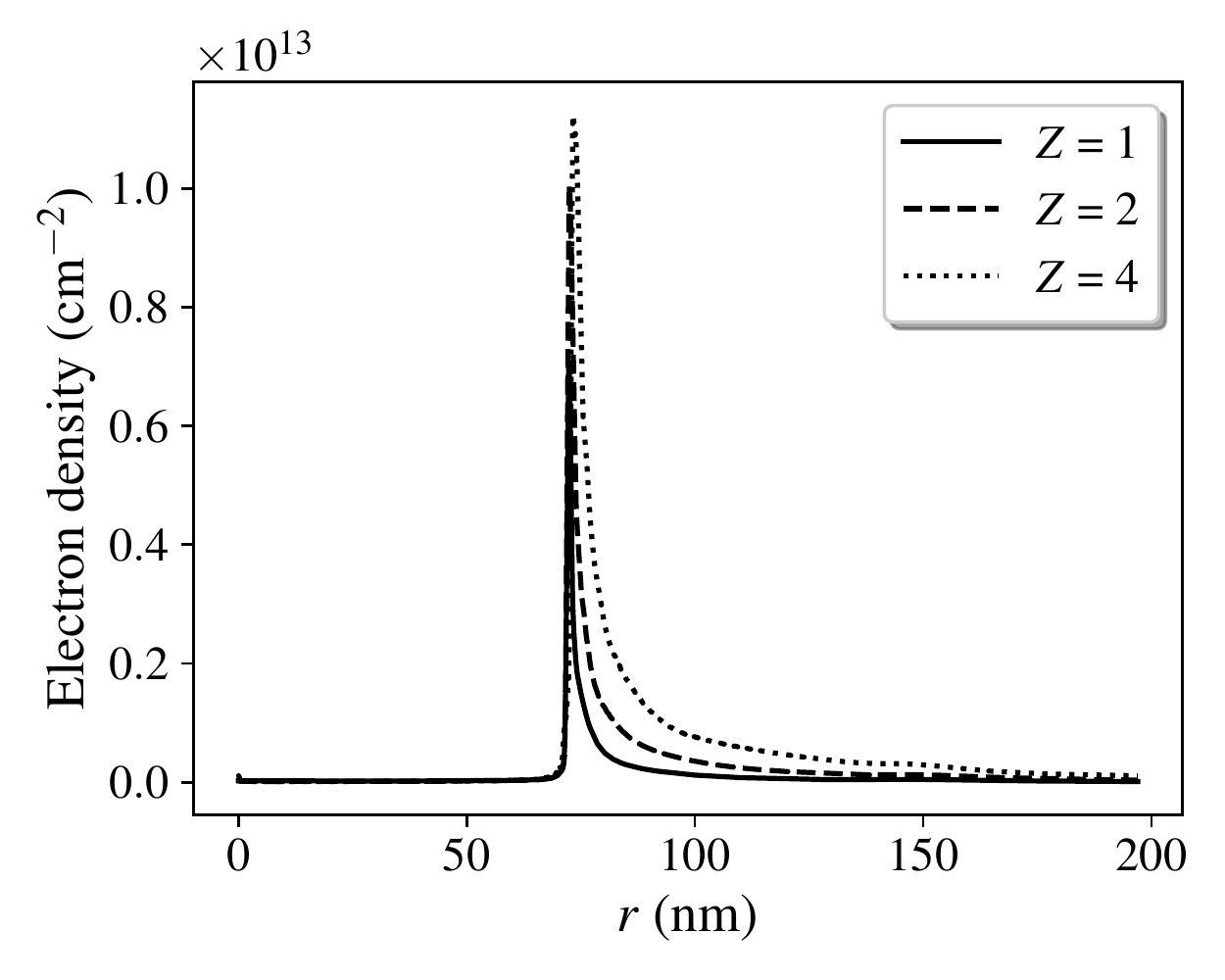}}
	\end{center}
	\caption{Dependence of the electron distribution and spatial density with the ion charge $Z$.}
	\label{fig:Z_dep}
\end{figure}

The numerical results for the electron distribution and electron spatial density as a function of the ion charge are displayed in Fig. \ref{fig:Z_dep}, for $v=0.001$. The results for different charges are qualitatively similar: there is a low energy peak in the electron distribution and the spatial distribution is a ring propagating radially. The main quantitative difference is that more energy states are excited for higher $Z$, resulting in a higher number of charge carriers. The total number of charge carriers, obtained by summing all energy and angular momentum contributions as $n = \sum_{kj_{z}} n_{kj_{z}}$ is given in Table \ref{tab:n} and shows a linear increase of the production rate. This can be attributed to the fact that for higher ion charges, more energy states become supercritical, resulting in additional excited tunneling pathways.

\begin{table}[t!]
	\begin{tabular}{ccc}
		\hline \hline
		Ion charge ($Z$) & Number of electrons ($n$) \\
		\hline
		1 & 22.7 \\
		2 & 52.1 \\
		4 & 101.1 \\
		\hline \hline
	\end{tabular}
	\caption{Total number of electrons as a function of the ion charge for $v=0.001$.}
	\label{tab:n}
\end{table}

%

\section{Conclusion \label{sec:conclu}}

In this work, the local field interband transitions induced by an ion passing through a graphene quantum dot have been studied theoretically. It is argued by looking at the spectrum and adiabatic conditions that the main process responsible for these transitions is APC, the same mechanism proposed for electron-pair production in low energy HIC. As a consequence, ion-bombarded graphene could serve as a testbed for QED physics in a more forgiving setting. Though APC eludes an experimental verification in HIC, it could be detected experimentally using our configuration. One possibility is to measure the radial induced current $I $. From the results of Section \ref{sec:field_induced}, we can roughly estimate that $I = Q/\Delta t \approx (2.0 \; \mbox{mA}) \cdot F$, where $F$ is the fraction of the circle where the current is measured and where we assumed that the number of charges generated is $n \approx 100$ while the time is estimated as $\Delta t \approx 10$ fs. According to this order of magnitude estimate, the induced current is in the several microamperes to low milliampere range, which is realistic for an experimental detection. A more elaborated and convincing approach would be to probe the electron distribution (Fig. \ref{fig:vel_dep}-(a) and \ref{fig:Z_dep}-(a))  using photoemission electron spectroscopy. However, this would be a challenging experiment, requiring a fine tuning of the ion spatial position and timing. More analysis are required to determine the feasibility of such an experiment.   

The theoretical investigations have been accomplished by introducing a Galerkin-like numerical scheme to solve the Dirac equation and by performing an extensive number of numerical calculations. The convergence of the numerical scheme has been demonstrated empirically. In principle, the numerical method could be used for other field configurations, such as the one provided by a nanotip, but it is restricted to azymuthally symmetric external potentials. A 2D Cartesian coordinate version of the numerical method would be possible if this condition is not fulfilled, although in this case, other numerical schemes could also be considered \cite{FillionGourdeau20121403,ANTOINE2017150}. 

Throughout the article, electron-electron interactions have been neglected. We expect that the electron distribution would not be modified significantly by these interactions, given that the obtained distribution are close to the thermal Fermi-Dirac distribution and the time scales involved. However, the effect on the spatial density could be more important and could induce a diffusion of the wave packet at larger times. This important topic is left for future work, along with the effect of different boundary conditions. Finally, the more complex process of carrier recombination could be investigated quantitatively because it could be used as a probe of electron/hole pair generation.

\begin{acknowledgments} 
The authors would like to acknowledge D. Gagnon for his participation in the coding of the numerical scheme and in many discussions on the article. We also thank H. Lu for his help with the numerical code. This research was enabled in part by support provided by Calcul Qu\'{e}bec (www.calculquebec.ca) and Compute Canada (www.computecanada.ca).
\end{acknowledgments}

\appendix

%

\section{Free solution of the Dirac equation}
\label{app:free_sol}

The free solution can then be found by setting $V(t,r)=0$. The Dirac equation then has the form
\begin{align}
\label{eq:free_dirac_eq}
i\partial_{t} 
\begin{bmatrix}
\psi_{1}(t,r)\\
\psi_{2}(t,r)
\end{bmatrix} 
=  i v_{F}
\begin{bmatrix}
0 & \partial_{r} + \frac{\mu_{2}}{r} \\
\partial_{r} - \frac{\mu_{1}}{r}
\end{bmatrix}
\begin{bmatrix}
\psi_{1}(t,r)\\
\psi_{2}(t,r)
\end{bmatrix},
\end{align}
where $\mu_{1} = j_{z}-1/2$ and $\mu_{2} = j_{z}+1/2$.
We make the following ansatz for positive and negative energy solution, respectively:
\begin{align}
\label{eq::ansatz_free1}
\psi(t,r) &= u(r)e^{-iEt}, \\
\psi(t,r) &= v(r)e^{iEt}.
\end{align}
We get the following coupled system of equations:
\begin{align}
Eu_{1}(r) &=  i v_{F}\left( \partial_{r}+ \frac{\mu_{2}}{r} \right) u_{2}(r), \\
Eu_{2}(r) &=  i v_{F}\left( \partial_{r}- \frac{\mu_{1}}{r} \right) u_{1}(r),
\end{align}
and
\begin{align}
-Ev_{1}(r) &=  i v_{F}\left( \partial_{r}+ \frac{\mu_{2}}{r} \right) v_{2}(r), \\
-Ev_{2}(r) &=  i v_{F}\left( \partial_{r}- \frac{\mu_{1}}{r} \right) v_{1}(r).
\end{align}
These equations can be decoupled and we get
\begin{align}
\left(\partial_{r}^{2} + \frac{1}{r}\partial_{r} - \frac{\mu_{2}^{2}}{r^{2}} + \frac{E^{2}}{v_{F}^{2}} \right)u_{2}(r), \\
u_{1}(r) =  i \frac{v_{F}}{E} \left( \partial_{r}+ \frac{\mu_{2}}{r} \right) u_{2}(r), \\
\left(\partial_{r}^{2} + \frac{1}{r}\partial_{r} - \frac{\mu_{2}^{2}}{r^{2}} + \frac{E^{2}}{v_{F}^{2}} \right)v_{2}(r), \\
v_{1}(r) = - i \frac{v_{F}}{E} \left( \partial_{r}+ \frac{\mu_{2}}{r} \right) v_{2}(r).
\end{align}
The second components $u_{2},v_{2}$ obeys the Bessel equation. Moreover, using the fact that the Bessel function derivative is given by
\begin{align}
\partial_{r} J_{m}(pr) &= -m\frac{J_{m}(pr)}{r} + kJ_{m-1}(pr), \\
\partial_{r} J_{m}(pr) &=  m\frac{J_{m}(pr)}{r} - kJ_{m+1}(pr),
\end{align}
and that $\mu_{1} = \mu_{2} -1$, we get the solutions
\begin{align}
\label{eq:ur}
u(r) &= N_{u}
\begin{bmatrix}
J_{\mu_{1}}(pr) \\
-i J_{\mu_{2}}(pr)
\end{bmatrix} \\
\label{eq:vr}
v(r) &= N_{v}
\begin{bmatrix}
J_{\mu_{1}}(pr) \\
i J_{\mu_{2}}(pr)
\end{bmatrix}
\end{align}
where $N_{u,v}$ are normalization constants to be determined and $p:= \sqrt{p_{x}^{2}+p_{y}^{2}} = E/v_{F}$. 

Now, boundary conditions at $r=0$ and $r=r_{\mathrm{max}}$ are taken into account. First, it is required  that solutions are not singular at the origin, omitting the contribution from Bessel functions of the second kind in the solution. Second, boxed boundary conditions are imposed by setting the first component to zero at the end of the domain, as $u_{1}(r_{\mathrm{max}}) = v_{1}(r_{\mathrm{max}}) = 0$. It is not possible to impose this condition to both spinor component at the same time and be consistent with the Dirac equation. Rather, the boxed boundary condition is set to one of the component (here the first) while the other component is evaluated from Eq. \eqref{eq:ti_u2}. Imposing this boundary condition, the momentum $p$ is quantized and takes the value $p_{k} = j_{\mu_{1},k}/r_{\mathrm{max}}$, for $k \in \mathbb{N}^{+}$ the principal quantum number and $j_{\mu,k}$ the zero of the Bessel function. For each value of $k$ corresponds an energy eigenstate.  


The normalization constant can now be evaluated by evaluating the product
\begin{align}
\langle u_{k} | u_{k} \rangle &= N_{k}^{2}\int_{0}^{r_{\mathrm{max}}} r \left[ J_{\mu_{1}}^{2}(p_{k} r) + J_{\mu_{2}}^{2}(p_{k}r)\right]dr.
\end{align}
This can be calculated using \cite{gradshteyn2014table}:
\begin{align}
\label{eq:int_bessel}
\int_{0}^{r_{\mathrm{max}}} rJ_{\mu}^{2}(pr) = \nonumber \\
 \frac{r_{\mathrm{max}}^{2}}{2} \left[J_{\mu}^{2}(pr_{\mathrm{max}}) - J_{\mu-1}(pr_{\mathrm{max}})J_{\mu+1}(pr_{\mathrm{max}}) \right],
\end{align}
yielding
\begin{align}
N_{k} = \frac{1}{r_{\mathrm{max}} \sqrt{J_{\mu_{1}+1}^{2}(j_{\mu_{1},k})}} .
\end{align}
Finally, it is verified that the positive and negative states are orthogonal on the domain by evaluating the product
\begin{align}
\langle u_{k}|v_{k} \rangle = N_{k}^{2} \int_{0}^{r_{\mathrm{max}}} r \left[ J_{\mu_{1}}^{2}(p_{k} r) - J_{\mu_{2}}^{2}(p_{k}r)\right]dr.
\end{align}
Again, using Eq. \eqref{eq:int_bessel}, it can be demonstrated that $\langle u_{k}|v_{k} \rangle = 0$.

\section{Numerical scheme for $j_{z} < 0$}
\label{app:neg_jz}

For $j_{z}<0$, the time-independent Dirac equation is written in a slightly different form where the spinor components are exchanged:  
\begin{align}
\label{eq:ti_u1_neg}
\left(\partial_{r}^{2} + \frac{1}{r}\partial_{r} - \frac{\mu_{2}^{2}}{r^{2}} + \frac{E^{2}}{v_{F}^{2}} \right)u_{2}(r)=0, \\
\label{eq:ti_u2_neg}
u_{1}(r) =  i \frac{v_{F}}{E} \hat{R}_{2} u_{2}(r). 
\end{align}
This choice for $j_{z}<0$ (and also the form for $j_{z}>0$) guarantees that singular terms of the form $1/r$ do not appear in the basis expansion nor in the energy functional, facilitating the numerical implementation. The energy functional for $j_{z}<0$ is given by
\begin{align}
\label{eq:functional_neg_jz}
\mathcal{E} &: = \lambda \int_{0}^{r_{\mathrm{max}}}rdr u_{2}^{\dagger}(r)u_{2}(r) \nonumber \\
&- \int_{0}^{r_{\mathrm{max}}}rdr u_{2}^{\dagger}(r) \left(\partial_{r}^{2} + \frac{1}{r}\partial_{r} - \frac{\mu_{2}^{2}}{r^{2}} \right)u_{2}(r).
\end{align}
where the eigenvalue is again $\lambda = \frac{E^{2}}{v_{F}^{2}}$. The polynomial basis is 
\begin{align}
\label{eq:basis_exp_neg_jz}
u_{2}(r) = r^{|\mu_{2}|} \sum_{i=1}^{N} a_{2,i} B_{2,i}(r).
\end{align}
Using the B-splines and substituting Eq. \eqref{eq:basis_exp_neg_jz} into Eq. \eqref{eq:functional_neg_jz}, we get another generalized eigenvalue problem where the explicit expression of the matrices are given in Appendix \ref{app:mat_tide}.
The solution of the eigenvalue problem then yields the second spinor component ($u_{2}$ and $v_{2}$, respectively). The other component $u_{1}$ is then 
\begin{align}
u_{1}(r) =   \sum_{i=1}^{N} a_{1,i} B_{1,i}(r), 
\end{align}
where 
\begin{align}
a_{1,i} &=  i \frac{v_{F}}{E}a_{2,i} ,\\
B_{1,i}(r) &= \hat{R}_{2} r^{|\mu_{2}|} B_{2,i}(r), \\
&=  r^{|\mu_{2}|} \partial_{r} B_{2,i}(r).
\end{align}

\section{Expression of matrices}

In this appendix, the explicit expression of matrices obtained in the discretization of the Dirac equation in cylindrical coordinates using the basis set expansion are given. 

\subsection{Time-independent Dirac equation}
\label{app:mat_tide}

The $A$ and $D$ matrices have non-zero entries when the basis functions have a common support. Therefore, if $\mathrm{supp}(B_{i}) \cap \mathrm{supp}(B_{j}) = \varnothing$ for $i,j \in [1,N]$, then $A_{ij} = D_{ij} = 0$, else, the non-zero entries are given by
\begin{itemize}
	\item For $j_{z} > 0$:
	\begin{align}
	A_{ij} &= \int_{0}^{r_{\mathrm{max}}} dr r^{2|\mu_{1}|+1} B_{1,i}(r) B_{1,j}(r), 
	\end{align}
	while
	\begin{align}
	D_{ij} &=  \int_{0}^{r_{\mathrm{max}}} dr r^{2|\mu_{1}|+1} [\partial_{r}B_{1,i}(r)][\partial_{r} B_{1,j}(r)]  .
	\end{align}
	\item For $j_{z} < 0$:	
	\begin{align}
	A_{ij} &= \int_{0}^{r_{\mathrm{max}}} dr r^{2|\mu_{2}|+1} B_{2,i}(r) B_{2,j}(r), 
	\end{align}
	while
	\begin{align}
	D_{ij} &= \int_{0}^{r_{\mathrm{max}}} dr r^{2|\mu_{2}|+1} [\partial_{r}B_{2,i}(r)][\partial_{r} B_{2,j}(r)]  .
	\end{align}	
\end{itemize}
An integration by parts of the functional $\mathcal{E}$ is required to obtain the expressions for $D_{ij}$.

\subsection{Time-dependent Dirac equation}
\label{app:mat_tdde}

As for the time-independent case, the $S$ and $C$ matrices have non-zero entries when the basis functions have a common support. Therefore, if $\mathrm{supp}(B_{i}) \cap \mathrm{supp}(B_{j}) = \varnothing$ for $i,j \in [1,N]$, then $S_{ij} = C_{ij} = 0$. In contrast to the time-dependent case, the non-zero entries are 2-by-2 matrices. They are given by
\begin{itemize}
	\item For $j_{z} > 0$:
	\begin{align}
	(S_{ij})_{11} &= \int_{0}^{r_{\mathrm{max}}} dr r^{2|\mu_{1}|+1} B_{1,i}(r) B_{1,j}(r), \\
	(S_{ij})_{22} &=  \int_{0}^{r_{\mathrm{max}}} dr r^{2|\mu_{1}|+1} [\partial_{r}B_{1,i}(r)][\partial_{r} B_{1,j}(r)] ,\\
	(S_{ij})_{12} &= (S_{ij})_{21} =0,
	\end{align}
	while
	\begin{align}
	(C_{ij})_{12} &=  i v_{F} \int_{0}^{r_{\mathrm{max}}} dr r^{2|\mu_{1}|+1} [\partial_{r}B_{1,i}(r)][\partial_{r} B_{1,j}(r)]  ,\\
	(C_{ij})_{21} &= (C_{ij})_{12}^{*}, \\
	(C_{ij})_{11} &= (C_{ij})_{22} =0,
	\end{align}
	and
	\begin{align}
	(P_{ij})_{11} &= \int_{0}^{r_{\mathrm{max}}} dr r^{2|\mu_{1}|+1} B_{1,i}(r) B_{1,j}(r)V(t,r), \\
	(P_{ij})_{22} &=  \int_{0}^{r_{\mathrm{max}}} dr r^{2|\mu_{1}|+1} [\partial_{r}B_{1,i}(r)][\partial_{r} B_{1,j}(r)]V(t,r) ,\\
	(P_{ij})_{12} &= (P_{ij})_{21} =0.
	\end{align}
	\item For $j_{z} < 0$:
	\begin{align}
	(S_{ij})_{11} &=  \int_{0}^{r_{\mathrm{max}}} dr r^{2|\mu_{2}|+1} [\partial_{r}B_{2,i}(r)][\partial_{r} B_{2,j}(r)] ,\\
	(S_{ij})_{22} &= \int_{0}^{r_{\mathrm{max}}} dr r^{2|\mu_{2}|+1} B_{2,i}(r) B_{2,j}(r), \\
	(S_{ij})_{12} &= (S_{ij})_{21} =0,
	\end{align}
	while
	\begin{align}
	(C_{ij})_{12} &= - i v_{F} \int_{0}^{r_{\mathrm{max}}} dr r^{2|\mu_{2}|+1} [\partial_{r}B_{2,i}(r)][\partial_{r} B_{2,j}(r)]  ,\\
	(C_{ij})_{21} &= (C_{ij})_{12}^{*}, \\
	(C_{ij})_{11} &= (C_{ij})_{22} =0,
	\end{align}
	and
	\begin{align}
	(P_{ij})_{11} &=  \int_{0}^{r_{\mathrm{max}}} dr r^{2|\mu_{2}|+1} [\partial_{r}B_{2,i}(r)][\partial_{r} B_{2,j}(r)]V(t,r) ,\\
	(P_{ij})_{22} &= \int_{0}^{r_{\mathrm{max}}} dr r^{2|\mu_{2}|+1} B_{2,i}(r) B_{2,j}(r)V(t,r), \\
	(P_{ij})_{12} &= (P_{ij})_{21} =0.
	\end{align}
\end{itemize}

\section{Convergence and numerical errors}
\label{app:num_conv}

In this appendix, the convergence of the numerical scheme is investigated, by performing simple test cases. Both the time-independent and time-dependent cases are considered.  

\subsection{Convergence of the time-independent scheme \label{app:num_conv_ti}}

The convergence of the time-independent scheme described in Section \ref{sec:time_ind} is studied by evaluating the free solution. These solutions given by Eqs. \eqref{eq:ur} and \eqref{eq:vr} are known analytically. The first test consists in verifying the spatial convergence as a function of the element size $h$ and the B-spline order. This is accomplished by evaluating the state with the lowest energy and by computing the numerical error, defined as the $L_{2}$-norm of the difference between the exact and the approximate solution $\epsilon = \left\lVert \psi_{\mathrm{appr}} - \psi_{\mathrm{exact}}\right\rVert_{L_{2}}$.

\begin{figure}[t!]
	\includegraphics[width=0.40\textwidth]{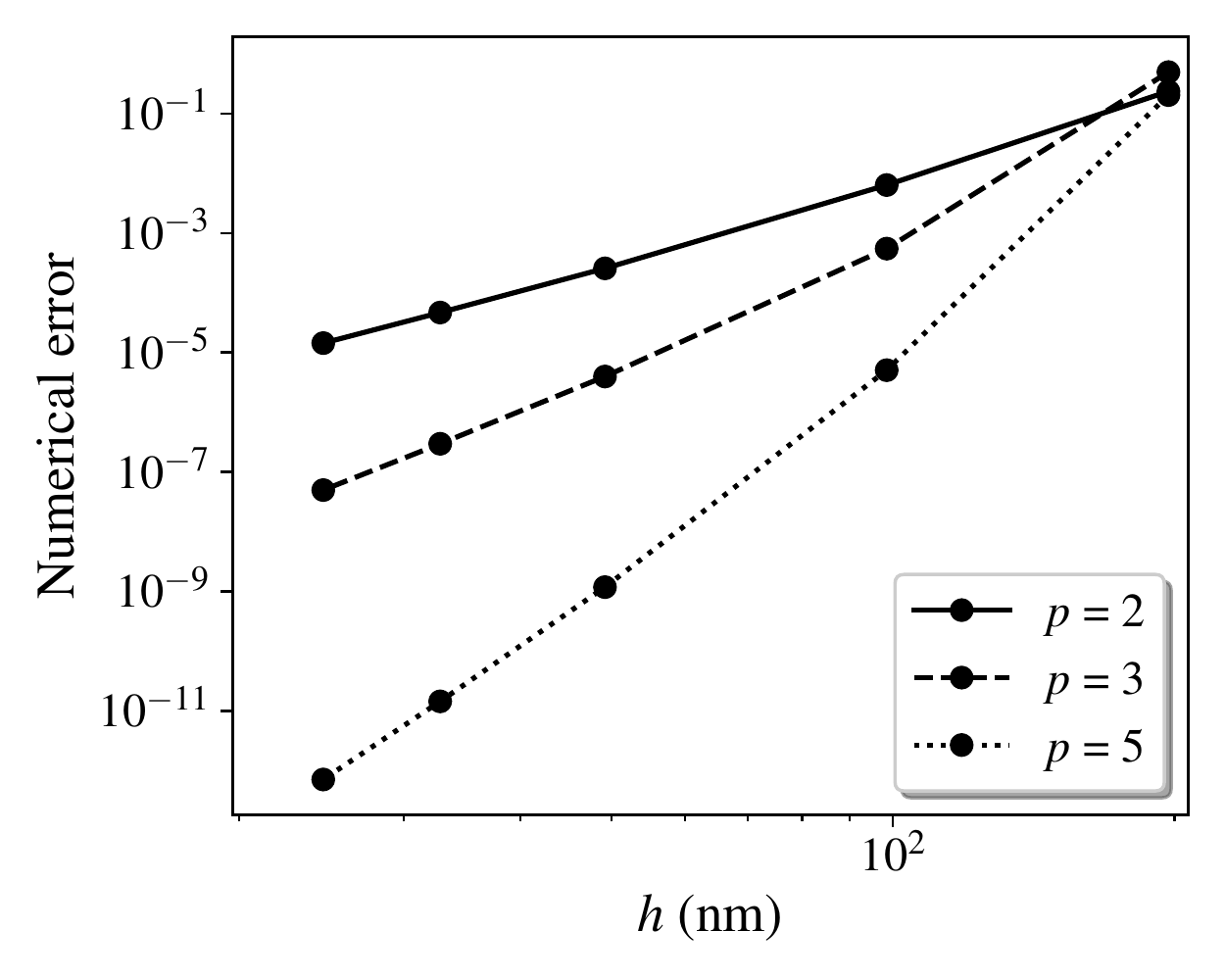}
	\caption{Numerical error as a function of element size $h$ and spline order $p$. }
	\label{fig:error_vs_h}
\end{figure}

The results are shown in Fig. \ref{fig:error_vs_h} for the numerical error while the order of convergence is given in Table \ref{tab:order}. The results demonstrate that the numerical scheme converges very rapidly as $h$ is reduced and the B-spline order is increased. As the computational time increases with the spline order because the matrices becomes less sparse, there is a compromise between accuracy and efficiency. In this article, we chose $p=4$ which converges rapidly but does not induce an important overhead.

\begin{table}[t!]
	\begin{tabular}{cc}
		\hline \hline
		B-spline order ($p$) & Order of convergence \\
		\hline
		2 & 4.66 \\
		3 & 7.66 \\
		5 & 12.60 \\
		\hline \hline
	\end{tabular}
	\caption{Order of spatial convergence for different B-spline order.}
	\label{tab:order}
\end{table}

\begin{figure}[t!]
	\includegraphics[width=0.40\textwidth]{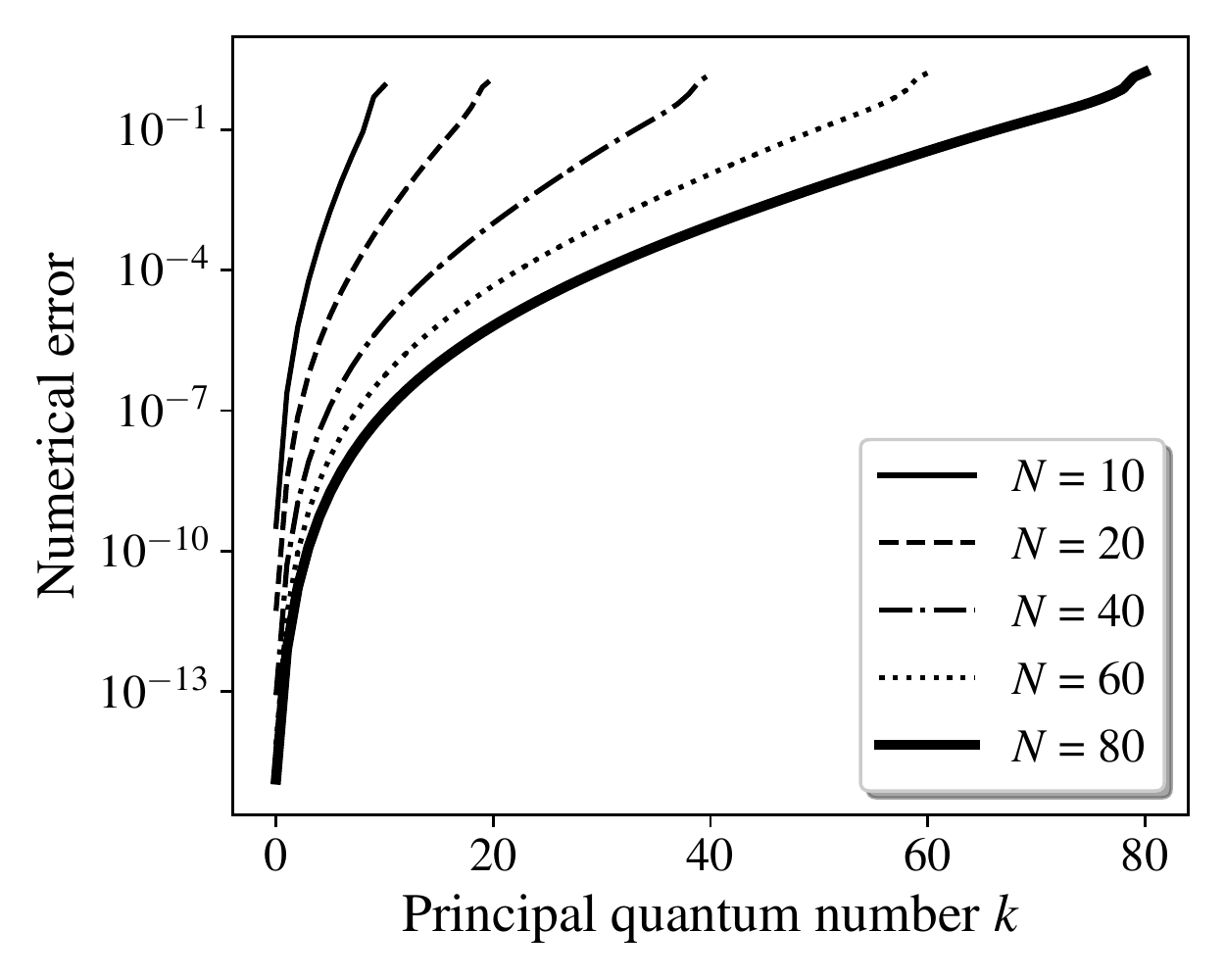}
	\caption{Numerical error as a function of eigenstate energies. The number on the abscissa represent the principal quantum number.}
	\label{fig:error_vs_eigen}
\end{figure}

In the second test, the accuracy of the numerical scheme is verified as function of eigenstate energies. All the eigenstates obtained from the numerical method are compared to the analytical solution. This is shown in Fig. \ref{fig:error_vs_eigen}, for many element lengths. These results demonstrate that the error increases with energy, up to a point where it reaches $O(1)$. This occurs because the period of spatial oscillations of the higher energy state wave functions approaches the grid resolution. According to these results, only the bottom half of all eigenstates are included in calculations in this article, allowing for accurate solutions.  

\subsection{Convergence of the time-dependent scheme \label{app:conv_td}}

\begin{figure}[t!]
	\includegraphics[width=0.40\textwidth]{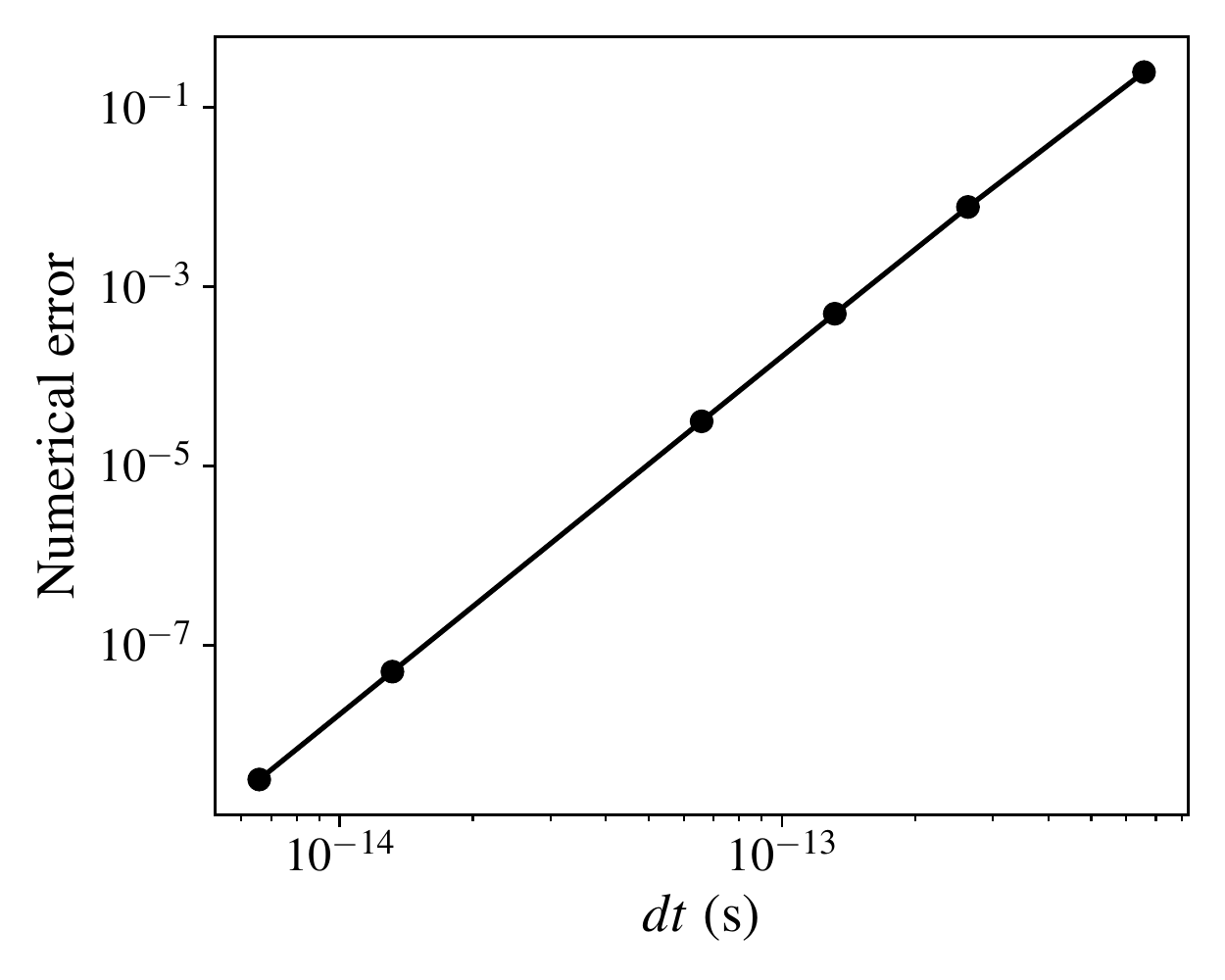}
	\caption{Numerical error of the time-dependent Galerkin scheme as a function of the time step $dt$. }
	\label{fig:error_vs_dt}
\end{figure}

The convergence of the time-dependent scheme is analyzed by propagating the positive energy ground state (with a principal quantum number $k=0$) from time $t_{i}=0$ s to $t_{f} = 6.58 \times 10^{-12}$ s and by varying the time step $dt$. The domain size is $r_{\mathrm{max}} = 1.97$ $\mu$m, the number of basis function is 40 and the spline order $p=4$. The error is evaluated using again the $L_{2}$-norm of the difference between the approximate solution and the exact solution. The analytical exact solution is given in Eqs. \eqref{eq::ansatz_free1} and \eqref{eq:ur}. The results are shown in Fig. \ref{fig:error_vs_dt}. These results demonstrate that the numerical scheme converge with time, with an order of convergence of 3.96. 

%
%
%
%



\bibliography{bibliography}

\end{document}